\documentclass[11pt]{article}

\usepackage{amsfonts,amsmath,amssymb,amsthm,mathrsfs}
\usepackage{adjustbox}
\usepackage{tikz}
\usepackage{ifthen}
\usepackage{fullpage}

\usepackage[ruled]{algorithm2e} 

\SetAlFnt{\small}
\SetAlCapFnt{\small}
\SetAlCapNameFnt{\small}
\SetAlCapHSkip{0pt}
\IncMargin{-\parindent}

\usepackage{booktabs} 
\usepackage{xspace}
\usepackage[pagebackref=true,hidelinks]{hyperref}
\usepackage{bbm}
\usepackage{enumitem}
\usepackage[numbers]{natbib}
\usepackage{multirow}
\usepackage{graphicx}
\usepackage{subfigure}
\usepackage{wrapfig}
\usepackage{cleveref}
\usepackage{nameref}
\usepackage{xcolor}

\newcommand*{\fullref}[1]{\hyperref[{#1}]{\cref*{#1} \nameref*{#1}}} 

\newtheorem{theorem}{Theorem}[section]
\newtheorem{proposition}{Proposition}[section]
\newtheorem{lemma}[theorem]{Lemma}
\newtheorem{corollary}[theorem]{Corollary}
\newtheorem{fact}[theorem]{Fact}
\newtheorem{remark}[theorem]{Remark}

\newtheorem{observation}[theorem]{Observation}

\newtheorem{definition}{Definition}[section]

\newcommand{\red}{\textcolor{red}}

\usepackage{mathtools}
\DeclarePairedDelimiter\abs{\lvert}{\rvert}%
\DeclarePairedDelimiter\norm{\lVert}{\rVert}%
\makeatletter
\let\oldabs\abs
\def\abs{\@ifstar{\oldabs}{\oldabs*}}
\let\oldnorm\norm
\def\norm{\@ifstar{\oldnorm}{\oldnorm*}}
\makeatother

\newcommand{\numberOfSlots}{k}
\newcommand{\numberOfAdvertisers}{n}

\newcommand{\numberofallocatedads}{m}
\newcommand{\setOfUsers}{U}

\newcommand{\separableModel}{position auction{}}
\newcommand{\simpleModel}[0]{Simple Model{}}
\newcommand{\ctr}[1]{\mathtt{CTR}_{#1}{}}
\newcommand{\ctru}[1]{\mathtt{CTR}^u_{#1}{}}
\newcommand{\userCTRVector}{\mathsf{\alpha}{}}
\newcommand{\userCTR}{\alpha}
\newcommand{\slotCTRVector}{\mathsf{\beta}}
\newcommand{\slotCTR}{\beta}

\newcommand{\betaname}[0]{slot-dependent CTR vector}
\newcommand{\alphasname}[0]{user-dependent CTR vectors}

\newcommand{\secondDef}[0]{value-stable for similarly-valued users with heterogeneous preferences}
\newcommand{\sw}[1]{SW(#1)}

\newcommand{\ipa}[0]{IPA}

\newcommand{\swalg}[0]{SW(\textsc{Alg})}
\newcommand{\swopt}[0]{\textsc{Unfair-Opt}}

\newcommand{\opt}[0]{\textsc{Unfair-Opt}}
\newcommand{\csum}[1]{C_{(#1)}}
\newcommand{\effv}[0]{\hat{v}}

\newcommand{\kipa}[1]{\mathsf{a}({#1})}
\newcommand{\kipaalloc}{\mathsf{a}}
\newcommand{\kIndivAlloc}{\mathsf{a}}
\newcommand{\kipah}[1]{\mathsf{a}^{(#1)}}
\newcommand{\mipa}[0]{M}

\newcommand{\separableMechanism}[1]{\mathcal{A}({#1})}

\newcommand{\weakvalueStable}[0]{\text{value stable}}

\newcommand{\kipaname}[0]{$k$-unit IPA}
\newcommand{\mipaname}[0]{Generalized IPA}
\newcommand{\sipaname}[0]{Generalized IPA}

\newcommand{\kunitsetting}[0]{$k$-unit setting}

\newcommand{\zerovector}[0]{\vec{0}}

\newcommand{\kpa}[0]{p}
\newcommand{\kpah}[1]{p^{(#1)}}
\newcommand{\mpa}[0]{P}

\newcommand{\vsum}[3]{\sum\limits_{i=#1}^{#2} v_i^{#3}}

\newcommand{\pa}[0]{PA}

\newcommand{\kpaname}[0]{$k$-unit PA}
\newcommand{\mpaname}[0]{Generalized PA}
\newcommand{\spaname}[0]{Generalized PA}
\newcommand{\kpaalloc}[0]{\mathsf{p}}

\newcommand{\hfair}[0]{ordered value stable}
\newcommand{\hfairness}[0]{ordered value stability}

\newcommand{\ofairness}[0]{ordered value stability}

\newcommand{\Ofairness}[0]{ordered value stability}

\newcommand{\remove}[1]{}
\newcommand{\etal}[0]{\textit{et al. }}
\newcommand{\f}[0]{f_{\ell}(\lambda)}

\begin{document}

\title{Individually-Fair Auctions for Multi-Slot Sponsored Search}
\author{
Shuchi Chawla \\ UT-Austin\\ {\tt shuchi@cs.utexas.edu} \and 
Rojin Rezvan \\ UT-Austin \\ {\tt rojinrezvan@utexas.edu} \and
Nathaniel Sauerberg \\ UT-Austin \\ {\tt njs@cs.utexas.edu} \and 
}
\date{}

\maketitle
\thispagestyle{empty}

\begin{abstract}
We design fair sponsored search auctions that achieve a near-optimal tradeoff between fairness and quality. Our work builds upon the model and auction design of Chawla and Jagadeesan \cite{CJ22}, who considered the special case of a single slot. We consider sponsored search settings with multiple slots and the standard model of click through rates that are multiplicatively separable into an advertiser-specific component and a slot-specific component. When similar users have similar advertiser-specific click through rates, our auctions achieve the same near-optimal tradeoff between fairness and quality as in \cite{CJ22}. When similar users can have different advertiser-specific preferences, we show that a preference-based fairness guarantee holds. Finally, we provide a computationally efficient algorithm for computing payments for our auctions as well as those in previous work, resolving another open direction from \cite{CJ22}. 
\end{abstract}
\newpage

\setcounter{page}{1}

\section{Introduction}\label{sec:intro}
We study the design of ad auctions under a fairness constraint. Fairness in the context of sponsored content has received considerable attention in recent years. It has been observed, for example, that ads on platforms such as Facebook and Google disproportionately target certain demographics, discriminating across users on the basis of race and gender. Furthermore, standard auction formats such as highest-bids-win can lead to discrimination even when the input to these algorithms, namely bids, CTRs, and relevance scores are themselves non-discriminatory.

\citet{CIJ20_Multi_Category_Fairness} initiated the study of optimal auction design under the constraint that the auction does not add any unfairness beyond what is already present in bids, and proposed a class of {\em proportional allocation} algorithms as a solution that achieves fairness while also providing an approximation to the optimal social welfare. In a followup work, \citet{CJ22} designed a class of {\em inverse proportional allocation} algorithms and showed that this class of mechanisms achieves an optimal tradeoff between social welfare and fairness. Both of these works focused on the simple case of a single item auction and left open the problem of designing a fair and efficient multi-slot position auction.

In this paper we extend the design of fair auctions from the single item setting to arbitrary position auction settings. We show that both the proportional allocation and inverse proportional allocation algorithms can be adapted to the setting of a position auction while inheriting their single-unit fairness properties as well as their approximation to social welfare. As in \cite{CIJ20_Multi_Category_Fairness, CJ22} our auctions provide fair solutions when the advertisers' bids are themselves non-discriminatory. Auctions for multi-slot settings must take into account both the advertisers' preferences over users as captured by per-click values, as well as the users' preferences over advertisers as captured by click through rates. We consider two different models for formalizing fairness in these settings. In the first, we consider differences of allocation across users that are close both in terms of the values advertisers assign to them as well as in terms of their own click through rates; we require that such users receive similar allocations. In the second setting, we consider pairs of users that are similarly qualified as per advertisers' values, but have different preferences (i.e. CTRs). In this case, while the users may receive different allocations, we require that allocations are suitably aligned with users' preferences. We elaborate on the details of these models below.
Finally, we address another open question in \cite{CIJ20_Multi_Category_Fairness, CJ22} and show how to efficiently compute supporting prices for both proportional and inverse proportional allocation.

\paragraph{Formalizing fairness across users.}
Consider two users Alice and Bob who are similar in most respects but differ in a sensitive demographic such as gender or race. Individual fairness then posits that Alice and Bob should see similar ad allocations. For example, it would be unfair to show more employment ads to Bob and more online retail ads to Alice. One potential source of unfairness in ad allocations is the use of discriminatory targeting by advertisers. However, empirical studies as well as theoretical analysis shows that unfairness in allocations can persist even in the absence of discriminatory targeting. The culprit is allocation algorithms that turn minor differences in advertisers' bids into large swings in allocation. Suppose, for example, that an employment agency places a slightly higher value on Bob than on Alice whereas an online retail store places a slightly higher value on Alice because of minor differences in the users' profiles. Then the highest-bid-wins auction would show entirely different ads to the two users.

To combat this problem, \citet{CJ22} formalize the notion of fairness in auctions as a ``value stability'' constraint. Informally speaking, value stability requires that whenever two users receive multiplicatively similar values from all advertisers (such as Alice and Bob in the example above) they must receive close allocations (as measured in terms of the $\ell_\infty$ distance between the respective probability distributions over the ad displayed). Previous work shows that while optimal auctions do not satisfy value stability, there are simple auction formats that do. In the {\em Proportional Allocation} (PA) mechanism, allocations are proportional to (some increasing function of) the advertisers' reported values. In the {\em Inverse Proportional Allocation} (IPA) mechanism, the unallocated amounts, i.e., one minus the probability of allocation, are inversely proportional to (some increasing function of) the advertisers' reported values. In both mechanisms, the allocation is a sufficiently smooth function of the advertisers' values and therefore satisfies some form of value stability. We mostly focus on the IPA mechanism in this paper as it provides better tradeoffs between fairness and welfare.

\paragraph{Multi-slot extensions.} As a simple extension of the single slot setting, consider a setting with $k$ slots, where each ad and each slot are equally likely to be clicked by the user, so the relative placement of ads in slots does not matter. In this case, one straightforward way to to extend the single-slot allocations is to simply multiply them by $k$; if this provides a valid allocation, the fairness and welfare guarantees follow immediately from the single-slot case. The problem is that some ads may receive a total allocation greater than $1$ and simply capping allocations at $1$ breaks the fairness guarantee. We propose a different extension of the IPA. As in the single slot case, we ensure that the unallocated amounts to advertisers are inversely proportional to (some function of) the reported values, subject to the total allocation equaling $k$. The fairness a.k.a. value stability of this extension follows easily from the single-slot special case. We further show that the social welfare approximation of multi-slot IPA matches its approximation for the single-item case by characterizing worst case instances for the approximation factor.

While the above discussion provides a complete story for the case of a multi-unit auction, in the case of online advertising, we also need to take click through rates into account. Throughout this paper, we assume that click through rates are multiplicatively separable into ad-specific and slot-specific components. In other words, the click through rate of an ad $i$ placed in slot $j$ is given by $\alpha_i\times\beta_j$ for some parameters $\alpha$ and $\beta$ specific to each user that are known to the platform/auctioneer. We further assume that all users weakly prefer earlier slots to later slots. Under these assumptions, we present an extension of the IPA to the ad auction setting that exactly maintains the social welfare guarantees of their single- and multi-unit counterparts. In particular, the social welfare approximation is independent of the number of slots. 

\paragraph{Fairness in the context of click through rates} is tricky to define, however. As before, we may assume that if two users are similarly qualified for all ads but differ in their sensitive attributes, then the two users receive multiplicatively similar per-click values from all advertisers. However, click through rates capture the users' own preferences and similar users may not have similar click through rates. What sort of fairness guarantees can we then provide?

We first show that differences in slot-specific CTRs do not impact fairness guarantees.\footnote{In fact, the allocations produced by our algorithms do not depend on the slot-specific CTRs, although the payments made by advertisers necessarily must.} In particular, two users with similar values and similar ad-specific CTRs $\alpha$ receive allocations that are close in $\ell_\infty$ distance. In particular, the probability of assigning any particular slot to any particular ad is additively close for the two users. In fact, this additive closeness holds also for the probability that any particular ad is assigned to slot $j$ or better for any $j$.

We then consider settings with similarly qualified users that have arbitrarily different ad-specific and slot-specific CTRs. Observe that in order to achieve any reasonable guarantee for social welfare, our allocation algorithms must take ad-specific CTRs into account. As a result, it is impossible to provide a value-stability guarantee in this setting while also providing an approximation to social welfare. Nevertheless, we show that a form of preference-aligned fairness holds. Specifically, let Alice and Bob be two users with multiplicatively similar values and let $\alpha$ and $\alpha'$ denote their ad-specific CTR vectors. Then we show that although the two users' allocations can be quite far from each other, Alice receives a higher allocation than Bob for precisely the ads that she is more likely to click on, and vice versa. Formally, if we sort the advertisers in decreasing order of the ratio $\alpha_i/\alpha'_i$, then for every $i$, the probability that Alice gets to see an ad with index $\le i$ is at least as large as Bob's probability of seeing the same set of ads.

\paragraph{Computing payments.} We conclude our study with a discussion of payments. It is easy to observe that both generalized IPA and generalized PA have monotone allocation rules in the advertisers' reported values. However, computing the supporting prices is not straightforward and was left open in previous work. Let $x_i(v_i)$ denote the net allocation (expected probability of click) to advertiser $i$ for a particular user, when the advertiser reports a per-click value of $v_i$. We show that $x_i(v_i)$ is a piecewise rational function with polynomially many pieces and that it is possible to compute the functional form of each piece in polynomial time. Computing payments using Myerson's lemma then boils down to computing polynomially many integrals over rational functions. 

\paragraph{Organization of the paper.} We present our extension of the IPA in Section~\ref{sec:ipa} and prove its social welfare and fairness guarantees for the setting of similarly qualified users with similar preferences. In Section~\ref{sec:ef} we discuss fairness for users that are similarly qualified but have different preferences. Section~\ref{sec:payments} presents our algorithm for computing payments. We extend our results to the PA in Section~\ref{sec:pa}. Most proofs are deferred to the appendix.

\subsection*{Related Work}

Journalism and empirical work have revealed the myriad ways in which existing ad auction systems lead to unfairness and discrimination \cite{10.1145/3359301, race,age,house,lbs967}. 
One approach to addressing these issues develops advertiser strategies for bidding in existing auction formats while ensuring statistical parity between groups \cite{GGMY20,10.1145/3351095.3375783}.

More related to our approach is theoretical work on designing auctions and, more generally, algorithms that guarantee fairness properties. These fairness properties typically differ in two dimensions: 1) whether they apply to individuals or only to groups as a whole, and 2) whether they enforce fairness by similarity of treatment or outcome, satisfaction of preferences (e.g., in the form of envy-freeness), or something bridging the two.



These notions of fairness grew out of the fair classification literature, where Dwork \etal \cite{10.1145/2090236.2090255} were the first to propose an individual fairness notion requiring agents who are similar under some task-specific metric to receive similar classifications. Dwork and Ilvento investigate in \cite{DI2018}  whether compositions of such classification algorithms that are fair in isolation maintain their fairness properties. 
Kim and Tschantz \cite{KKRY2019} introduce individual preference-informed fairness by augmenting this notion of individual fairness with envy-freeness, allowing the allocations of similar users to differ in accordance with their preferences. 
Similarly, Zafar \etal in \cite{parity_to_preferences} develop notions of preference-informed group fairness by allowing deviations from parity in treatment and impact if the deviations are envy-free. 




Our work employs and expands upon a model of individual fairness in sponsored search first developed by \citet{CIJ20_Multi_Category_Fairness} and based on the multi-category fairness work of \citet{DI2018}. An alternate model, also based on \cite{DI2018}, was presented by \citet{watts2021fairness}, albeit in a Bayesian setting. A main difference between our work and  \cite{watts2021fairness} is that we study the design of auctions that achieve an optimal tradeoff between fairness and welfare, whereas \cite{watts2021fairness} analyzes the fairness and welfare of two specific mechanisms. Another relevant work is that of \citet{essaidi2021symmetries} who study the fairness-welfare tradeoff in a Bayesian setting. \cite{essaidi2021symmetries} draws a connection between individual fairness in this context and multi-item auctions  with an item symmetry constraint, giving simple mechanisms that achieve a constant-approximation to the revenue-optimal fair mechanism.


There is also some recent work on group-fair ad auctions, such as \cite{yuan2021fair}, which shows that constraints on advertiser behavior which enforce group fairness notations can actually increase the profit of the platform. 
In a Bayesian setting, \cite{celli2021learning} augments generalized second price auctions with fair division schemes to achieve good social welfare guarantees while satisfying envy-freeness properties among advertiser groups. 

As far as we know, ours is the first work addressing fairness specifically in the positional auctions setting where different users have different click through rates.

\section{Models and Definitions}\label{sec:model}
We consider the following stylized model for online advertising auctions. Let $\setOfUsers$ be the set of users, $\numberOfAdvertisers$ the number of advertisers, and $\numberOfSlots$ the number of slots. We use index $u$ for users, $i$ for advertisers and $j$ for slots. At each point in time, a user $u\in\setOfUsers$ arrives. Each advertiser $i\in [n]$ bids a per-click value $v_i^u$ on that user. This is the value the advertiser receives if the user clicks on their ad. Let $\ctru{i,j}$ denote the click through rate of advertiser $i$ in slot $j$, that is, the probability that the user $u$ will click on the ad $i$ if it is placed in slot $j$. 

A truthful auction decides which ads to display in each of the $\numberOfSlots$ slots. The auction receives the vector $v=(v^u_1,\ldots, v^u_{\numberOfAdvertisers})$ as well as the click through rates $\ctru{.}$ and returns an allocation matrix $\kipa{v}=[\kIndivAlloc_{ij}]_{i\in [\numberOfAdvertisers], j\in[\numberOfSlots]}$ where $\kIndivAlloc_{ij}$ denotes the probability that ad $i$ is displayed in slot $j$.\footnote{We require $\sum_i \kIndivAlloc_{ij} = 1$ for all $j$ and $\sum_j \kIndivAlloc_{ij} \le 1$ for all $i$. Every matrix $\kipa{\cdot{}}$ satisfying these matching constraints can be expressed as a distribution over deterministic assignments of ads to slots.} We omit the superscript $u$ whenever it is clear from the context that we are discussing a certain user.


\paragraph{Truthfulness.} Given an allocation $\kipa{v}$ (where the user $u$ is implicit), advertiser $i$ receives a net allocation (expected number of clicks) of $\sum_j \ctru{i,j} \kIndivAlloc_{ij}$ and a net expected value of $v_i \cdot \sum_j \ctru{i,j} \kIndivAlloc_{ij}$ from the allocation. To ensure truthfulness, there should exist a supporting pricing function $p_i(v)$ for every advertiser $i$ such that bidding truthfully maximizes the advertiser's net expected utility. For such a payment function to exist, it is sufficient and necessary that the allocation probability $\sum_j \ctru{i,j} \kIndivAlloc_{ij}$ is monotone non-decreasing in the per-click value $v_i$. All of the mechanisms we discuss in this paper satisfy monotonicity. In Section~\ref{sec:payments} we discuss how to compute supporting payments efficiently.


\paragraph{Separable click through rates.} Throughout this paper we assume that the click through rates $\ctru{i,j}$ are multiplicatively separable into an advertiser-specific component and a slot-specific component. This is a standard model (see, for example, \cite{10.1145/1134707.1134708}).

\begin{definition}[Separable Click Through Rates] Click through rates are \textnormal{separable} if, for every user $u$, there exists a advertiser dependent vector $\userCTRVector_u=(\userCTR_1,\ldots,\userCTR_{\numberOfAdvertisers})$ and a slot dependent vector $\slotCTRVector_u=(\slotCTR_1,\ldots,\slotCTR_{\numberOfSlots})$ in which $\userCTR_1,\ldots,\userCTR_{\numberOfAdvertisers}>0$ and $1\geq\slotCTR_1\geq \slotCTR_2\geq \ldots \geq \slotCTR_{\numberOfSlots}\geq 0$ such that $\ctru{i,j}=\userCTR_i\slotCTR_j$ for all $i\in[\numberOfAdvertisers]$ and $j\in[\numberOfSlots]$. 
\end{definition}

All of the mechanisms we design and analyze in this paper are based on the vector of values normalized by the ad-specific CTR of the user; We call these the "effective values" of the advertisers:
\begin{definition}[Effective Value] The 
 effective value of advertiser $i$ is given by $\effv_i = v_i\userCTR_i$.
\end{definition}

We call the above model of online advertising auctions with separable CTRs the {\bf Position Auction Setting}.

\paragraph{Prior-free design.} As in previous works, the mechanisms we design and analyze in this paper are prior-free, meaning that the allocation to a user does not depend on the distribution of users or advertisers' value vectors or the history of users already served. Besides the well-documented benefits of prior-free mechanism design, in the context of fairness we get the added benefit that fairness guarantees hold for all users that are served by the mechanism regardless of whether or not the auctioneer's model accounts for them.

\begin{definition}[Scale-Free]\label{def:scale-free}
    A mechanism is scale-free if it has the property that multiplying the input values by a uniform constant does not change the resulting allocation. 
\end{definition}


\subsection{Social Welfare}

The goal of this work, as in \cite{CJ22, CIJ20_Multi_Category_Fairness}, is to achieve a tradeoff between fairness and social welfare for the mechanisms we design. 
The social welfare of an allocation $\kipa{v}$ is defined to be the sum of all of the advertisers' net expected values:
\[\sw{\kipa{v}}=\sum_{i\in [\numberOfAdvertisers],j\in[\numberOfSlots]} v_i \ctru{i,j} \kIndivAlloc_{i,j}\]

We compare this social welfare to the maximum achievable by any feasible allocation. When click through rates are separable, the maximum social welfare is achieved by the allocation that assigns  advertisers to slots in decreasing order of $\hat{v_i}$, the effective values. We call the allocation sorted by effective values the {\sc Unfair-Opt} and also use the same term to denote the social welfare of this allocation. 

Formally, if $\pi$ is the order of advertisers where $\hat{v}_{\pi_1}\geq \hat{v}_{\pi_2}\geq\ldots\geq \hat{v}_{\pi_n}$, then the (unfair) optimal social welfare is given by:
\[\swopt(v,\userCTRVector,\slotCTRVector)=\sum_{j=1}^k \userCTR_{\pi_j}v_{\pi_j} \slotCTR_j\]


Since it is generally impossible to achieve optimal social welfare and fairness simultaneously, we look for mechanisms that  guarantee our fairness notions while giving a good approximation to the optimal social welfare.


\begin{definition}[Social Welfare Approximation]
We say mechanism $\separableMechanism{\cdot{}}$ achieves an $\eta$-approximation to social welfare for $\eta\leq 1$, if for all instances $(v,\userCTRVector, \slotCTRVector)$, we have $\sw{\separableMechanism{v,\userCTRVector,\slotCTRVector}}\geq \eta \cdot{} \swopt(v,\userCTRVector,\slotCTRVector)$.
\end{definition}

\subsection{Fairness}
\cite{CJ22} formalized fairness in ad auctions as a value stability condition based on the notion of individual fairness. Individual fairness requires that the auction assign similar allocations to similar users. \cite{CJ22} defined similarity between two users on the basis of closeness between the value vectors assigned to them by the advertisers. Informally speaking, if two users receive similar values from all advertisers, then they should also receive similar allocations. In order for the definition to be scale-free with respect to values, similarity between values is defined in multiplicative terms.

In the context of a single item auction, allocations are probability vectors. Similarity in allocations is therefore defined based on some notion of distance between probability vectors. \cite{CJ22} formalized similarity in terms of the $\ell_{\infty}$ distance between the probability vectors whereas \cite{CIJ20_Multi_Category_Fairness} used total variation or $\ell_{1}$ distance. We state the value stability definition from \cite{CJ22} below.


\begin{definition}[Definition 2.1 from \cite{CJ22}, Value Stability] \label{def:original-value-stable}
An allocation mechanism $\kipa{\cdot}$ is value stable with respect to function $f : [1,\infty] \rightarrow [0,1]$ if the following condition is satisfied for every pair of value vectors $v$ and $v'$:
$$|\kipaalloc_i(v) - \kipaalloc_i(v')| \leq f(\lambda) \text{ for all } i\in[n], \text{ where } \lambda=\max\limits_{i\in[n]} \left( \max\left\{\frac{v_i}{v'_i}, \frac{v'_i}{v_i}\right\}\right).$$
\end{definition}

In this definition, the function $f$, called the value stability constraint, governs the strength of the value stability condition. We assume $f$ to be non-decreasing, with $f(0)=0$ and $f(\infty)=1$. Following \cite{CJ22}, we focus on the family of constraints $f_{\ell}(\lambda)=1-\lambda^{-2\ell}$. \cite{CJ22} argue that this family of stability constraints captures the entire spectrum of possible fairness conditions in the context of allocation algorithms. 

In order to extend these fairness definitions to the position auctions setting, we need to extend the notion of closeness in allocations to multi-dimensional allocation matrices $\mipa$ as well as extend the notion of closeness in values to click through rates. 

Let us consider the latter issue first. A straightforward manner of extending closeness over value vectors to the separable setting is to require that two similar users are assigned similar values, as well as have similar click through rates. But this notion of closeness is too restrictive. 
Values capture how advertisers perceive users as potential customers; whereas click through rates capture how users perceive the relevance of ads to their needs and how users behave in perusing ads on a search page. Two users that are similarly qualified for a set of ads may nevertheless exhibit very different behavior in responding to ads on a search page. 
Ideally the fairness guarantees an allocation algorithm provides should hinge only on the closeness between values $v_i$ and not on the closeness between click through rates $\ctr{i,j}$.
However, in order to obtain good social welfare, allocations necessarily need to depend on the advertiser specific click through rates $\alpha_i$. We accordingly define closeness between users in terms of their effective values $\alpha_iv_i$ (while ignoring dissimilarity in slot specific CTRs, $\beta$).
In Section~\ref{sec:ef} we extend our fairness definitions and guarantees to settings where closeness is defined only in terms of the values $v_i$, ignoring dissimilarity in $\alpha$ and $\beta$. 

Let us now consider closeness over probability matrices. We consider three notions. The first is $\ell_\infty$ distance, the maximum difference of allocations in any one entry $(i,j)$ of the corresponding matrices. 


\begin{definition}[Value Stability for Position Auctions]\label{def:weakly-value-stable}
An allocation mechanism $\separableMechanism{\cdot}$ is $\weakvalueStable$ with respect to function $f: [1,\infty]\rightarrow [0,1]$ if the following condition is satisfied for every set of value and CTR vectors $v$, $v'$, $\userCTRVector$, $\userCTRVector'$ and $\slotCTRVector$:
$$|\mipa_{i,j}-\mipa'_{i,j}|\leq 2f_{\ell}(\lambda) \text{ for all } i\in[n], j\in[k] \text{ where } \lambda \text{ is defined as } \max_{i\in[n]} \left( \max\left\{\frac{\alpha_iv_i}{\alpha'_iv'_i}, \frac{\alpha'_iv'_i}{\alpha_iv_i}\right\}\right)$$
and
$\mipa= \separableMechanism{v,\userCTRVector,\slotCTRVector}$ and $\mipa'= \separableMechanism{v',\userCTRVector',\slotCTRVector}$.
\end{definition}

Suppose, as an example, for a particular advertiser $i$, user $u$ has an allocation of $a = (.1, .1, .1, .1)$. Consider two possible allocation vectors for some $v$ close to $u$: $a' = (.15, .15, .15, .15)$ and $a'' = (.15, .05, .15, .05)$. In some sense, allocation $a'$ is much more unfair than $a''$ because in $a'$ the entry-wise differences from $a$ compound while in $a''$ they offset each other. Weak value stability cannot distinguish these two cases because it is concerned only with the absolute differences. Our next definition, \Ofairness\ is intended to allow $a''$ but not $a'$.

To do this, we bound the absolute differences in the total allocation of an advertiser across all columns, weighted by a vector $h_{i,j}$. This vector represents the utility the first user receives from seeing advertisement $i$ in slot $j$. Since we assume the slots are in decreasing order of salience, this should be weakly decreasing in $j$. 


\begin{definition}[Ordered Value Stability for Position Auctions] 
 An allocation mechanism $\separableMechanism{\cdot}$ is \hfair\ with respect to function $f: [1,\infty]\rightarrow [0,1]$ if the following condition is satisfied for every set of value and CTR vectors $v$, $v'$, $\userCTRVector$, $\userCTRVector'$ and $\slotCTRVector$, as well as for any advertiser $i$ and any decreasing vector $h_i$ with $1 \geq h_{i,1} \geq \ldots \geq h_{i, k} \geq 0$:
 
$$\abs{ \sum_{j =1}^k h_{i,j} \left(\mipa_{i,j}-\mipa'_{i,j}\right) } \leq f_{\ell}(\lambda) \text{ where } \lambda \text{ is defined as } \max_{i\in[n]} \left( \max\left\{\frac{\alpha_iv_i}{\alpha'_iv'_i}, \frac{\alpha'_iv'_i}{\alpha_iv_i}\right\}\right)$$
where
$\mipa= \separableMechanism{v,\userCTRVector,\slotCTRVector}$ and $\mipa'= \separableMechanism{v',\userCTRVector',\slotCTRVector}$.
\end{definition}

The previous two definitions are concerned only with a single advertiser. In some instances, however, there are meaningful subsets of advertisers and bounding the differences of the allocations each advertiser individually may not be sufficient to ensure fairness overall. For example, if there are several different ads giving information about registering to vote, the total volume of voter registration ads a user sees is more important from a fairness perspective than the amount they see any particular voter registration ad. 
Therefore, the last notion we consider is a combination of $\ell_1$ and $\ell_\infty$ distance: we consider, for any subset of advertisers, the total variation distance between the allocations of these advertisers to one slot, and bound the maximum over all slots of this distance.

\begin{definition}[Total Variation Value Stability for Position Auctions] \label{def:TV-value-stability} 
    A mechanism $\separableMechanism{\cdot{}}$  with satisfies total variation value stability with respect to a function $f: [1,\infty] \rightarrow [0,1]$ if the following condition is satisfied for every set of value and CTR vectors $v$, $v'$, $\userCTRVector$, $\userCTRVector'$ and $\slotCTRVector$, as well as every subset of advertisers $S \subseteq [n]$ and for every column $j$:
    $$\abs{\sum_{s \in S} \separableMechanism{\effv}_{s,j} - \sum_{s \in S}  \separableMechanism{\effv}_{s,j}} \leq f(\lambda)
    \text{ where } \lambda \text{ is defined as } \max_{i\in[n]} \left( \max\left\{\frac{\alpha_iv_i}{\alpha'_iv'_i}, \frac{\alpha'_iv'_i}{\alpha_iv_i}\right\}\right)$$
and where
$\mipa= \separableMechanism{v,\userCTRVector,\slotCTRVector}$ and $\mipa'= \separableMechanism{v',\userCTRVector',\slotCTRVector}$.
\end{definition}

\section{Inverse Proportional Allocation}\label{sec:ipa}
In this section, we present a generalization of the mechanism first introduced in \cite{CJ22} as \ipa\ to the \separableModel\ setting. We show that the generalization retains a constant approximation to the optimal social welfare and an appropriate generalization of the value stability condition. In \Cref{IPA:algorithm} we describe the generalization of the mechanism from $k=1$ to general $k$. In \Cref{IPA:fairness} we show that two different value stability conditions hold and in \Cref{IPA:sw} we show that the exact same guarantee in \cite{CJ22} holds for the generalization as well. All proofs in this section are deferred to \Cref{sec:appendix-ipa}.

\subsection{Generalized IPA}\label{IPA:algorithm}
In \cite{CJ22}, \ipa\ was presented as a mechanism for the single item auction. 
An interpretation of this mechanism is as follows: start with an infeasible allocation of $1$ unit to each advertiser (for a total allocation of $n$) and then gradually decrease the allocations until the total allocation reaches $1$. The rate of this decrease is determined by a function $g$ of the reported values. The IPA with parameter $\ell$ uses $g(x)=x^{-\ell}$. \cite{CJ22} also presents an algorithmic interpretation of the mechanism. The following is the generalization of this mechanism to the \separableModel\ setting.

First, as a warm-up, we generalize \ipa\ to a special case of the \separableModel\ setting where $\slotCTRVector = \overrightarrow{1}$. Our algorithm allocates a total of $k$ units to the advertisers, with each advertiser receiving an allocation $a_i\in [0,1]$ such that $\sum_i a_i=k$.

We follow the same intuition as for the case of $k=1$. The mechanism first allocates $1$ to each advertiser, then decreases the allocations until the total allocation reaches $\numberOfSlots$ rather than $1$. See \Cref{sec:appendix-ipa} for an algorithmic interpretation of this mechanism. Note that setting $\numberOfSlots=1$ gives the exact same mechanism as in \cite{CJ22}. 
Algorithm \ref{alg:k-unit-IPA} is scale free and produces allocations that are non-decreasing in $k$. Furthermore, the allocation to advertiser $i$, namely $a_i$, is non-decreasing in $\effv_i$ and non-increasing in $\effv_{-i}$.


\remove{
This is not technically a solution to the setting of the \simpleModel, which requires an $n \times k$ matrix. However, simply using $\frac{1}{k}\kipaalloc$ as every column vector in the matrix gives a solution to the \simpleModel\ setting, while the vector formulation will be useful as we use the $k-$unit algorithm (\Cref{alg:k-unit-IPA}) as a subroutine in the next section. In \Cref{lemma:k-unit-ipa-realizable}, we show that there the allocation produced by this mechanism is always feasible, meaning that there exists a distribution over matchings from advertisers to slots, for which the total probability that advertiser $i$ is allocated a slot is equal to $\kIndivAlloc_i(v)$.
}

We now extend the $k$-unit setting to the \separableModel\ setting. The resulting allocation algorithm is called Generalized IPA. The algorithm assigns to every slot $j$ a distribution over advertisers given by the difference in the $j$-unit and $j-1$-unit allocations produced by $k$-unit IPA. 

\vspace{0.5 cm}
\begin{algorithm}[H]
 \textbf{Input: }Vector $v$ of non-negative advertiser bids for user $u$; CTRs $\userCTR_1, \cdots, \userCTR_n$ and $\slotCTR_1, \cdots, \slotCTR_{\numberOfSlots}$; number of slots $\numberOfSlots$; function $g:\mathbb{R}^{\geq 0} \rightarrow (0,\infty]$ with $g(0)= \infty$ and $\lim_{x\rightarrow \infty}g(x)=0$\;
 \For {$h \in [k]$}{Set $\kipah{h}\leftarrow$ the output of the IPA k-unit algorithm 
 on input $(v, \userCTRVector, h, g)$}
 \For{$j \in [\numberOfSlots]$}{Set $\mipa_{\cdot,j} = \kipah{j} - \kipah{j-1}$ }
 \Return \mipa
 \caption{\mipaname}\label{alg:IPA-separable}
\end{algorithm}
\vspace{0.5 cm}

Note that the generalized IPA algorithm is scale-free and independent of $\slotCTRVector$. 

\paragraph{Feasibility.}We observe that the allocation produced by the generalized IPA algorithm is feasible. That is, there exists a distribution over matchings from advertisers to slots, for which the total probability that advertiser $i$ is allocated a slot is equal to $\mipa$. 

\begin{fact}\label{thm:seperable-ipa-feasible}
Let $M$ be the $n\times k$ matrix output by arbitrary run of the generalized IPA algorithm on input $(v, \userCTRVector, \slotCTRVector, k, g)$. Let $\mathcal{M}$ be the set of matchings from the set of advertisers $([\numberOfAdvertisers])$ to the set of slots $([\numberOfSlots])$. There exists a probability distribution $\mathcal{P}$ over $\mathcal{M}$ such that for all advertisers $i$ and slots $j$, the probability that $i$ is assigned to slot $j$ is equal to the probability prescribed by the matrix $M$, that is $\Pr_{m\sim \mathcal{P}}[ (i, j) \in m] = M_{i,j}$. 
\end{fact}

\subsection{Fairness}\label{IPA:fairness}
We now prove the value stability of the \mipaname\ mechanism. 

\begin{theorem}\label{thm:sep-IPA-value-stable}
    The \mipaname\ mechanism with parameter $\ell > 0$ and for any number of advertisers $\numberOfAdvertisers$ is \weakvalueStable\ with respect to any function $f$ satisfying $f(\lambda) \geq f_\ell(\lambda) = 1 - \lambda^{-2\ell}$ for all $\lambda \in [1, \infty)$, as in \Cref{def:weakly-value-stable}.
\end{theorem}

Our proof has two parts. First, give a bound on the deviation between allocations given by the \kipaname\ mechanism to similar users. Then, we use the bound to show that \mipaname\ achieves value stability. 

\begin{lemma}\label{lemma:kipa-value-stable}
    For the \kipaname\ mechanism with parameter $\ell$ run on any $k$ and any bid vectors $v$ and $v'$ with $\lambda = \max_{i \in [n]} \{\hat{v}_i / \hat{v}_i' , \hat{v}_i'/ \hat{v}_i\}$, for all indices $i$, $|\kipaalloc_i(v) - \kipaalloc_i(v') | \leq f_\ell(\lambda)$.
\end{lemma}

Next, we show that Generalized IPA satisfies ordered value stability. 
\begin{theorem}\label{thm:ipa-ordered-value-stability}
    \mipaname\ with parameter $\ell$ satisfies \hfairness\ with respect to $f_\ell(\lambda)$. That is, for every set of value and CTR vectors $v$, $v'$, $\userCTRVector$, $\userCTRVector'$ and $\slotCTRVector$, as well as for any advertiser $i$ and any decreasing vector $h$ with 
     $1 \geq h_{1} \geq \ldots \geq h_{k} \geq 0$:
 
$$\abs{ \sum_{j =1}^k h_{j} \left(\mipa_{i,j}-\mipa'_{i,j}\right) } \leq f_{\ell}(\lambda) \text{ where } \lambda \text{ is defined as } \max_{i\in[n]} \left( \max\left\{\frac{\alpha_iv_i}{\alpha'_iv'_i}, \frac{\alpha'_iv'_i}{\alpha_iv_i}\right\}\right)$$
where
$\mipa= \separableMechanism{v,\userCTRVector,\slotCTRVector}$ and $\mipa'= \separableMechanism{v',\userCTRVector',\slotCTRVector}$.
\end{theorem}

\subsection{Social Welfare}\label{IPA:sw}

We now show that \mipaname\ achieves a good approximation to the optimal social welfare $\opt$. 

\begin{theorem}\label{thm:ipa-separable-sw}
    The IPA algorithm for the separable case, \Cref{alg:IPA-separable}, run with parameter $\ell > 0$ and any number of advertisers $\numberOfAdvertisers$ achieves a $\left(1-\frac{\ell^{\ell}}{(1+\ell)^{\ell+1}}\right)$-approximation the social welfare of the unfair optimum.
\end{theorem}

To do so, we first show an approximation result for the special case of $\Vec{\beta} = 1$, the $k-$unit algorithm.

\begin{lemma}\label{lemma:k-unit-IPA-SW}
    The IPA algorithm for the $k-$unit case, \Cref{alg:k-unit-IPA}, run with parameter $\ell$ and any number of advertisers $\numberOfAdvertisers$ achieves a $\left(1-\frac{\ell^{\ell}}{(1+\ell)^{\ell+1}}\right)$-approximation to the social welfare of the unfair optimum. 
\end{lemma}
We use \Cref{lemma:k-unit-IPA-SW} and extend definition of Generalized IPA allocation vector based on $k-$unit vectors to show \Cref{thm:ipa-separable-sw}. 
The approximation factor is $\frac{3}{4}$ at $\ell=1$ and as $\ell\rightarrow\infty$, the approximation factor goes to $1$.

\begin{remark}
The approximation factor in \Cref{lemma:k-unit-IPA-SW} is tight for IPA mechanism.
\end{remark}
\begin{proof}
Consider the following example. Fix a user $u$ and let the bidding vector of the advertisers be:
$$\mathlarger{(\underbrace{1,\ldots,1}_{k},\overbrace{\epsilon,\ldots,\epsilon}^{n-k})}$$
where $1>\epsilon=\frac{-5k+\sqrt{25k^2-16(n-k)\frac{k^2}{n-k}-4-4(n-k)}}{8(n-k}>0$. Let $\ell=1$ and $n>2k$. We get:
$$\swalg=k(1-\frac{n-k}{(n-k)\epsilon^{-1}+k})+(n-k)\epsilon(1-(n-k)\frac{\epsilon^{-1}}{(n-k)\epsilon^{-1}+k}), \quad \swopt=k$$
For the aforementioned value of $\epsilon$, we will have $\frac{\swalg}{\swopt}=\frac{3}{4}$. Note that this example fits the maxima point we found in the proof of \Cref{lemma:k-unit-IPA-SW}.
\end{proof}

\section{Fairness for users with different preferences}\label{sec:ef}
So far we have assumed that similar users are similar in all aspects -- the values advertisers assign to them as well as the rates at which the users click on different ads. However, these two sets of parameters are asymmetric. Values capture advertisers' preferences over users whereas CTRs capture users' preferences over advertisers. We will now distinguish between similarity in {\em qualification} (i.e. values) from similarity in {\em user preferences} (i.e. CTR). 

A myopic viewpoint might suggest that two users that are similarly qualified should be treated similarly by the auction no matter their preferences. However, this is fundamentally at odds with the objective of maximizing the social welfare\footnote{Social welfare is a misnomer in this context, as it does not take into account the benefit or value users derive from viewing the ad.} a.k.a. the collective value of the advertisers, as the latter are contingent upon clicks. Consequently, the outcome of the auction cannot be completely independent of user preferences and we look towards a notion of fairness that is appropriately preference aligned. 

To motivate our definitions, consider the following example. We have two users Alice and Bob, two advertisers A and B, and a single slot to display an ad. The users look identical to the advertisers: A places a value of \$1 on a click from either user and B places a value of \$10 from either click. However the users behave differently when they view ads. Bob clicks both ads with certainty. Alice clicks A's ad with certainty but B's ad with probability only 1\%. The platform should clearly display ad A for Alice and ad B for Bob. Although these outcomes are different, both users are happy: Bob is essentially indifferent between A and B, while Alice greatly prefers A. In this case, any differences in allocation are aligned with user preferences.

Can we always expect this to be the case? Formally, consider a single slot auction with $n$ advertisers, and two users with identical value vectors $v=v'$. Let $\kipaalloc$ and $\kipaalloc'$ denote their respective allocation vectors. Can we ensure that any allocation mass that is moved between advertisers in $\kipaalloc'$ relative to $\kipaalloc$ is moved from low CTR advertisers to high CTR advertisers? 

Unfortunately, we cannot ensure this property while also maintaining a reasonable approximation for social welfare. To see this, consider the above example with Alice and Bob once again and suppose that Bob's CTR for advertiser B changes to 20\%. In order to obtain a good social welfare, the auction must continue to display ad B for Bob. However, now Bob gets to see much more of ad B and much less of ad A than Alice even though he greatly prefers ad A to ad B. The key observation here is that the allocation mass in B's allocation shifts to an advertiser with high {\em relative} CTR, when measured relative to the CTRs of Alice. 

Motivated by this example, we propose the following new preference-aligned definition of fairness for identically valued users. Underlying this definition is a relative ordering of advertisers for two users $u$ and $v$ with advertiser specific CTR vectors $\alpha_u = (\alpha^u_1, \cdots, \alpha^u_n)$ and $\alpha_v = (\alpha^v_1, \cdots, \alpha^v_n)$. We will assume that advertisers are ordered in (weakly) decreasing order of the ratio $\alpha^v_i/\alpha^u_i$, and require that allocation mass for user $v$ is shifted from advertisers that appear later in the ordering to those that appear earlier in the ordering.

\begin{definition}[Value Stability for Identically-Valued Users with Heterogeneous
 Preferences]\label{def:EF-separable}
An allocation mechanism $\separableMechanism{\cdot}$ is value-stable for identical users with heterogeneous preferences if for every pair of users with identical value vectors $v$; CTR vectors $\userCTRVector$, $\userCTRVector'$, $\slotCTRVector$, and $\slotCTRVector'$; any ordering over advertisers that is weakly decreasing in $\userCTRVector/\userCTRVector'$; and for every advertiser $i\in[n]$ and slot $j \in [k]$:
$$\sum_{t=1}^i\sum_{s=1}^j \mipa_{t,s} \geq \sum_{t=1}^i\sum_{s=1}^j \mipa'_{t,s}, \quad \text{ where } \mipa =\separableMechanism{v,\userCTRVector, \slotCTRVector} \text{ and } \mipa' =\separableMechanism{v,\userCTRVector', \slotCTRVector'}.
$$
\end{definition}

\textbf{Similar users:}
The above definition extends in a straightforward manner to pairs of users that are similarly rather than identically qualified, and again have different preferences over advertisers as expressed through CTRs. Once again we require that allocation mass shifts from advertisers with low relative CTR to those with higher relative CTR, but we allow for additive errors in allocation that grow with the dissimilarity in the users' values.


\begin{definition}[Value Stability for Similarly-Valued Users with Heterogeneous
 Preferences]\label{def:PIF-weak-separable}
 An allocation mechanism $\separableMechanism{\cdot}$ is value-stable for users with heterogeneous preferences with respect to function $f_{\ell} : [1, \infty] \rightarrow [0,1]$  if for every pair of users with value vectors $v$ and $v'$; CTR vectors $\userCTRVector$, $\userCTRVector'$, $\slotCTRVector$, and $\slotCTRVector'$; any ordering over advertisers that is weakly decreasing in $\userCTRVector/\userCTRVector'$; and for every advertiser $i\in[n]$ and slot $j \in [k]$:
$$\sum_{t=1}^i\sum_{s=1}^j \mipa_{t,s} \geq \sum_{t=1}^i\sum_{s=1}^j \mipa'_{t,s} - i f_\ell(\lambda), \quad \text{ where } \mipa =\separableMechanism{v,\userCTRVector, \slotCTRVector} \text{ and } \mipa' =\separableMechanism{v',\userCTRVector', \slotCTRVector'},
$$
where $\lambda=\max\limits_{i\in[n]}\left\{\max\left\{\frac{v_i}{v'_i}, \frac{v'_i}{v_i}\right\}\right\}$.
\end{definition}
Comparing \Cref{def:EF-separable} and \Cref{def:PIF-weak-separable}, note that if $v=v'$ then 
$\lambda=1$ and, as discussed in \cite{CJ22}, a proper $f $ function has the property of $f(1)=0$. Therefore, \Cref{def:EF-separable} is exactly \Cref{def:PIF-weak-separable} in the special case of $v=v'$.

\remove{\begin{definition}\label{def:PIF-strong-separable}
[Strong Preference Informed Fairness for $\separableModel$] An allocation mechanism $\separableMechanism{\cdot}$ with slot-dependent vector $\slotCTRVector$ is $\strongpreferenceInformedFair$ with respect to function $f_{\ell} : [1, \infty] \rightarrow [0,1]$ if the following condition is satisfied for every pair of users with value vectors $v$ and $v'$ and \alphasname\ $\userCTRVector, \userCTRVector'$ and every advertiser $i \in [\numberOfAdvertisers]$ and slot $j \in [\numberOfAdvertisers]$:

$$\sum_{t=1}^i \kipaalloc_{\pi_t} \geq \sum_{t=1}^i \kipaalloc'_{\pi_t} - f_{\ell}(\lambda)$$ 
$$\text{ where } \mipa =\separableMechanism{v,\userCTRVector, \slotCTRVector},\hspace{2pt} \mipa' =\separableMechanism{v,\userCTRVector', \slotCTRVector},\hspace{2pt} \kipaalloc_t = \sum_{s=1}^{j} \mipa_{t,s}, \text{ and } \kipaalloc'_t = \sum_{s=1}^{j} \mipa'_{t,s},$$
and where $\pi$ is a permutation on advertisers for which $\frac{\userCTR_{\pi_1}}{\userCTR'_{\pi_1}}\geq \ldots \geq \frac{\userCTR_{\pi_{\numberOfAdvertisers}}}{\userCTR'_{\pi_{\numberOfAdvertisers}}}$.
\end{definition}}


\subsection{Fairness of IPA and PA for heterogeneous users}
We show that both the \sipaname\ and \spaname\ mechanisms satisfy \Cref{def:EF-separable} and more generally \Cref{def:PIF-weak-separable}. 

To begin, we show that any mechanism for the $k$-unit case satisfying certain mild conditions also satisfies \Cref{def:EF-separable}. Both k-unit IPA and k-unit PA satisfy these conditions and hence are value-stable for identically qualified users with heterogeneous preferences.

\begin{lemma}\label{EF_k-unit-algs}
	Let $\kipa{v}$ be a scale-free $k$-unit allocation algorithm such that $\kipaalloc_i(v)$ is weakly increasing in $v_i$. Suppose further that for all $t \neq i$, $\kipaalloc_i(v)$ is weakly decreasing in $v_{t}$. Then $\kipa{v}$ satisfies \Cref{def:EF-separable}.
\end{lemma}

\begin{proof}
	Fix $i$ and scale $\alpha'$ so that $\alpha_i = \alpha_i'$. Since the advertisers are sorted, we now know that for all $t < i$, $\alpha_t \geq \alpha_t'$ and for all $t > i$, $\alpha_t \leq \alpha_t'$.
	
	We proceed by two cases and then use a transitivity argument to show the theorem holds in general. 
	
	Consider the case where for all $t \leq i$, $\alpha_t = \alpha_t'$. Then $\alpha v \begin{cases}
		= \alpha' v \mbox{ for all } t \leq i \\		
		\leq \alpha' v \mbox{ for all } t > i \\
	\end{cases}.$

	Therefore, since the allocation $a_t$ is weakly decreasing in $v_s$ for all $s \neq t$, we have that for all $t \leq i$, $a(\alpha v) \geq a(\alpha' v)$. Hence, $\sum_{t = 1}^i a_t(\alpha v) \geq \sum_{t = 1}^i a_t(\alpha' v)$, as desired. 
	
	Now, consider the case where for all $t \geq i$, $\alpha_t = \alpha_t'$. Then $\alpha v \begin{cases}
	 \geq \alpha' v \mbox{ for all } t < i \\		
	 = \alpha' v \mbox{ for all } t \geq i \\
	\end{cases}.$

	Therefore, since the allocation $a_t$ is weakly decreasing in $v_s$ for all $s \neq t$, we have that for all $t > i$, $a(\alpha v) \leq a(\alpha' v)$ and hence $\sum_{t = i+1}^n a_t(\alpha v) \leq \sum_{t = i+1}^n a_t(\alpha' v)$. But $\sum_{t = 1}^i a_t(\alpha v) = k - \sum_{t = i+1}^n a_t(\alpha v)$ and likewise $\sum_{t = 1}^i a_t(\alpha' v) = k - \sum_{t = i+1}^n a_t(\alpha' v)$. Therefore, $\sum_{t = i+1}^n a_t(\alpha v) \leq \sum_{t = i+1}^n a_t(\alpha' v)$ implies $\sum_{t = 1}^i a_t(\alpha v) \geq \sum_{t = 1}^i a_t(\alpha' v)$, as desired.
	
	We now argue that the theorem holds in general. Let $\alpha''_t := \begin{cases} 
		\alpha_t \mbox{ if } t\leq i\\
		\alpha'_t \mbox{ if } t > i
	\end{cases}$.
	By the first case, $\sum_{t = 1}^i a_t(\alpha v) \geq \sum_{t = 1}^i a_t(\alpha'' v)$, and by the second case $\sum_{t = 1}^i a_t(\alpha'' v) \geq \sum_{t = 1}^i a_t(\alpha' v)$. Hence, $\sum_{t = 1}^i a_t(\alpha v) \geq \sum_{t = 1}^i a_t(\alpha' v)$, as desired. 
\end{proof}


\begin{corollary}\label{cor:kipa-ef}
	The \kipaname\ and \kpaname\ mechanisms satisfy \Cref{def:EF-separable}.
\end{corollary}

\remove{\begin{proof} 
    The \kipaname\ mechanism is clearly scale-free. Further, $\kIndivAlloc_i$ is clearly monotone non-decreasing in $v_i$ and monotone non-increasing in $v_t$ for all $t\neq i$.
\end{proof}

\begin{corollary}\label{cor:kpa-ef}
	The \kpaname\ mechanism satisfies \Cref{def:EF-k-unit}.
\end{corollary}

\begin{proof} 
    The \kpaname\ mechanism is clearly scale-free. Further, $\kIndivAlloc_i$ is clearly monotone non-decreasing in $v_i$ and monotone non-increasing in $v_t$ for all $t\neq i$.
\end{proof}

Next, we show that the \sipaname\ and \spaname\ mechanisms satisfy their envy-freeness definition, \Cref{def:EF-separable}.
}

Because our generalized mechanisms are defined in terms of telescoping differences of the $k$-unit allocations, \Cref{value-stable-identical-users} follows directly from \Cref{cor:kipa-ef}. 

\begin{theorem}\label{value-stable-identical-users}
    The \mipaname\ and \mpaname\ mechanisms satisfy \Cref{def:EF-separable}.
\end{theorem}

Next, we show \sipaname\ and \spaname\ are \secondDef. The only thing changing from \Cref{def:EF-separable} to \Cref{def:PIF-weak-separable} is that we need to keep track of small changes between the two allocations, which leads to the following theorem. The proof is deferred to \Cref{sec:appendix-ef-proofs}.

\begin{theorem}\label{thm:ipa_weak_piif}
    The \mipaname\ and \mpaname\ mechanisms $\separableMechanism{\cdot}$ with parameter $\ell$ are \secondDef. 
\end{theorem}

\section{Computing payments}\label{sec:payments}
In this section we develop an algorithm for computing supporting payments for the generalized IPA and generalized PA allocation rules. Our main observation is that the allocation functions of IPA and PA are piecewise rational functions with polynomially many pieces where each piece can be computed in polynomial time. With these pieces in hand, and using Myerson's lemma, computing payments amounts to computing polynomially many integrals of rational functions.

We focus on the generalized IPA; the argument for generalized PA is similar. Formally, for a fixed and implicit user $u$, and a fixed and implicit advertiser $i$, let $x_i(v)$ denote the net allocation to the advertiser, a.k.a. the expected number of clicks the advertiser receives from the user. If the user is assigned allocation $M=\separableMechanism{v,\userCTRVector,\slotCTRVector}$ then we have $x_i(v)=\sum_j M_{i,j}\alpha_i\beta_j$. Let $a^{(j)}$ denote the cumulative allocation to the user in the first $j$ slots as in the description of Algorithm 2 and recall that $M_{i,j} = a^{(j)}_i-a^{(j-1)}_i$. Accordingly we get:
\begin{align}
    x_i(v) & = \alpha_i \sum_j a^{(j)}_i(\beta_j-\beta_{j+1}) \label{eqn:linear-combo}
\end{align}
In other words, $x_i(v)$ is a linear combination of the functions $a^{(j)}_i(v)$.

We will now argue that for all $i,j$, the function $a^{(j)}_i(v)$, as defined in Algorithm 1, is piecewise rational in $v_i$. Consider the following equivalent formulation of Algorithm 1. Given the values $v_1, \cdots, v_n$, ad-specific CTRs $\alpha_1, \alpha_2, \cdots, \alpha_n$, and decreasing function $g$, we find a parameter $t$ such that 
\begin{align}
    \sum_{i'} \min(1,t \cdot g(\alpha_{i'}v_{i'})) & = n-j \label{eqn:simple-alg1}
\end{align}
The allocation $a^{(j)}_i$ is then given by $1-\min(1,t \cdot g(\alpha_{i}v_{i}))$.

Suppose without loss of generality that $i$ receives a non-zero allocation at value $v_i$ (otherwise $a^{(j)}_i$ is trivially piecewise rational at values $\le v_i$). We can then rewrite Equation~\eqref{eqn:simple-alg1} as:
\begin{align}
    t \cdot g(\alpha_{i}v_{i}) + \sum_{i'\ne i} \min(1,t \cdot g(\alpha_{i'}v_{i'})) & = n-j \label{eqn2:simple-alg1}
\end{align}

Now, the expression $\sum_{i'\ne i} \min(1,t g(\alpha_{i'}v_{i'}))$ is independent of $v_i$ and piecewise linear in $t$ with at most $n$ pieces. Given the values $v_{-i}$ and CTRs $\alpha_{-i}$, we can efficiently compute the linear pieces in this function. Substituting any particular linear piece with $t$ in the range $[t_1,t_2]$ in Equation~\eqref{eqn2:simple-alg1} then gives us an equation of the following form with appropriate parameters $x$ and $y$:
\[t \cdot g(\alpha_{i}v_{i}) + xt = y\]
leading to the solution
\[a^{(j)}_i(v_i) = 1-g(\alpha_{i}v_{i}) \cdot \frac{y}{g(\alpha_{i}v_{i})+x} \quad \text{for } v_i\in \left[\frac{1}{\alpha_i} g^{-1}\left(\frac{y-x t_2}{t_2}\right), \frac{1}{\alpha_i} g^{-1}\left(\frac{y-x t_1}{t_1}\right) \right].\]
Observe that the RHS in the above equation is a rational function as the function $g$ in the definition of IPA is also rational.

Summarizing, we first compute the piecewise rational form of the function $a^{(j)}_i(v_i)$ for all slots $j$. Each of these functions has at most $n$ pieces. We then use Equation~\eqref{eqn:linear-combo} to express $x_i(v_i)$ as a piecewise rational function with at most $nk$ pieces. Finally, we use Myerson's lemma and compute per-impression payments as
\[p_i(v_i) = v_i x_i(v_i) - \int_{z=0}^{v_i} x_i(z) \, dz.\]

\section{Proportional Allocation}\label{sec:pa}
In this section, we present a generalization of the mechanism first introduced in \cite{CIJ20_Multi_Category_Fairness} as Proportional Allocation (PA) to the position auction setting. We show that the generalization retains the same approximation ratio to the optimal social welfare and an appropriate generalization of the total variation value stability condition. This is a stronger fairness guarantee than that of \mipaname, but comes at the cost of a weaker approximation to the optimal social welfare. For a detailed discussion of the trade-offs between the single-unit versions these methods, see \cite{CJ22}. All proofs in this section are deferred to \Cref{sec:appendix-pa}.


\subsection{Generalized PA}

In contrast to \ipa, \pa\ can be thought of as initially assigning each advertiser an allocation of $0$ and then increasing the allocations in proportion to (some function of) the bid amounts until the total allocation reaches $1$. \cite{CIJ20_Multi_Category_Fairness} analyzes this mechanism for the single unit case. In particular, they prove value stability with respect to the total variation distance on the allocations, rather than with respect to the $\ell^\infty$ distance as with \ipa. 
However, in exchange, the social welfare approximation achieved by \pa\ degrades as the number of advertisers increases. 

Just like the previous section, we start with a warm-up case in which we consider a special case of \separableModel where $\slotCTRVector = \overrightarrow{1}$. For this case, we will attempt to allocate proportionally, assigning $k\cdot{}\frac{g(v_i)}{\sum_t g(v_t)}$ to each bidder $i$. If this allocation is more than $1$ for any advertiser, we cap their allocation at $1$ and divide the additional mass proportionally among the remaining advertisers. See \Cref{PA-algorithm} in \Cref{sec:appendix-pa} for an algorithmic interpretation of this mechanism. 
Note that the function $g$ in this mechanism is different than the one in \Cref{sec:ipa}, as it is a continuous, super-additive and increasing function. 

The extension of this algorithm to the \separableModel case is similar to the extension we saw in \Cref{sec:ipa} for IPA, and works as follows:

\vspace{0.5 cm}
\begin{algorithm}[H]
 \textbf{Input: }Vector $v$ of non-negative advertiser bids for user $u$; CTRs $\userCTR_1, \cdots, \userCTR_n$ and $\slotCTR_1, \cdots, \slotCTR_{\numberOfSlots}$; number of slots $\numberOfSlots$; function $g:\mathbb{R}^{\geq 0} \rightarrow [0,\infty]$ with $g$ a continuous, super-additive, increasing function and $g(0)=0$\;
 \For {$h \in [k]$}{Set $\kpah{h}\leftarrow$ the output of the PA k-unit algorithm 
 on input $(v, \userCTRVector, h, g)$ }
 \For{$j \in [\numberOfSlots]$}{Set $\mpa_{\cdot,j} = \kpah{j} - \kpah{j-1}$ }
 \Return \mpa
 \caption{\mpaname}\label{alg:PA-separable}\label{alg:PA-General}
\end{algorithm}
\vspace{0.5 cm}

Observe that Generalized PA is scale-free, independent of $\beta$, and produces feasible allocations by essentially the same argument as in \Cref{thm:seperable-ipa-feasible}.

\subsection{Fairness}

First, we prove the fairness guarantees of our mechanism. We begin by showing the total variation value stability of PA, which as we've discussed is the main advantage of PA over IPA. 

\begin{theorem}\label{thm:PA_fairness}
    The \mpaname\ mechanism with parameter $g(x) = x^\ell$ satisfies \fullref{def:TV-value-stability} with respect to $f_\ell(\lambda)$. That is, for all pairs of effective value vectors $\effv, \effv'$, subsets of advertisers $S \subseteq [n]$, and slots $j$, 
    $$ \abs{\sum_{s \in S} P_{s,j}(\effv) - \sum_{s \in S} P_{s,j}(\effv')} \leq 2 f_\ell(\lambda)$$
\end{theorem}

The proof of \Cref{thm:PA_fairness} uses the following key lemma, which shows a similar property holds for \kpaname\ mechanism. 

\begin{lemma}\label{lemma:k-unit-PA-fairness}\label{lemma:k-unit-pa-tv-value-stability}
    The \kpaname\ mechanism with parameter $g(x) = x ^\ell$ satisfies the property that, for all pairs of effective value vectors $\effv, \effv'$ and subsets of advertisers $S \subseteq [n]$, 
    $$\abs{\sum_{s \in S} a_s(\effv) - \sum_{s \in S} a_s(\effv')}  \leq \frac{\lambda^\ell - 1}{\lambda^\ell + 1} \leq f_\ell(\lambda).$$
\end{lemma}

We now show that \mpaname\ also satisfies the same \hfairness\ property as IPA. The proof is essentially identical as the proof of \Cref{thm:ipa-ordered-value-stability} except in that it uses the total variation value stability of PA instead the value stability of IPA. For the full proof, see \Cref{sec:appendix-pa}.

\begin{theorem}\label{thm:pa-ordered-value-stability}
    \mpaname\ with parameter $\ell$ satisfies \hfairness\ with respect to $f_\ell(\lambda)$. That is, for every set of value and CTR vectors $v$, $v'$, $\userCTRVector$, $\userCTRVector'$ and $\slotCTRVector$, as well as for any advertiser $i$ and any decreasing vector $h$ with 
     $1 \geq h_{1} \geq \ldots \geq h_{k} \geq 0$:
 
$$\abs{ \sum_{j =1}^k h_{j} \left(\mpa_{i,j}-\mpa'_{i,j}\right) } \leq f_{\ell}(\lambda) \text{ where } \lambda \text{ is defined as } \max_{i\in[n]} \left( \max\left\{\frac{\alpha_iv_i}{\alpha'_iv'_i}, \frac{\alpha'_iv'_i}{\alpha_iv_i}\right\}\right)$$
where
$\mpa= \separableMechanism{v,\userCTRVector,\slotCTRVector}$ and $\mpa'= \separableMechanism{v',\userCTRVector',\slotCTRVector}$.
\end{theorem}

    
\subsection{Social Welfare}
Finally, we give our guarantee on the social welfare approximation ratio achieved by \mpaname\ relative to \opt. The proof relies on a lemma showing the same approximation result for the special case of $\Vec{\beta} = 1$, \kpaname. 


\begin{theorem}\label{thm:PA_SW}
    The \mpaname\ mechanism with parameter $\ell$ achieves a $\left( \frac{n-k}{n}(n-k)^{-1/\ell} + 1/n \right)$-approximation to the optimal social welfare for any instance with $n$ advertisers and $k$ slots. 
\end{theorem}


\begin{lemma}\label{lemma:k-unit-PA-SW}
    The \kpaname\ subroutine with parameter $\ell$ achieves a $\left( \frac{n-k}{n}(n-k)^{-1/\ell} + 1/n \right)$-approximation to the optimal social welfare for any instance with $n$ advertisers and $k$ slots. 
\end{lemma}

\bibliographystyle{plainnat}
\bibliography{ref}

\appendix

\section{Deferred Proofs from \Cref{sec:ipa}}\label{sec:appendix-ipa}
\subsection{Algorithm}
Below is the algorithmic description of \separableModel in the case of $\Vec{\beta}=1$:

\vspace{0.5 cm}
\begin{algorithm}[H]
 \textbf{Input: }Vector $v$ of non-negative advertiser bids for user $u$; ad-specific CTRs $\userCTR_1, \cdots, \userCTR_n$; number of slots $\numberOfSlots$; function $g:\mathbb{R}^{\geq 0} \rightarrow (0,\infty]$ with $g(0)= \infty$ and $\lim_{x\rightarrow \infty}g(x)=0$\;
 \textbf{Initialization:}
 Determine effective values, $\effv_i=v_i\alpha_i$ for all $i$\;
 WLOG assume $\effv_1 \geq \ldots\geq \effv_{\numberOfAdvertisers}$\;
 \If{$k=0$}
 {\Return $\kipa{v}=\overrightarrow{0}$}
 \If{$\effv_1\leq 0$}{
 Set $\kipaalloc_i = \frac{\numberOfSlots}{\numberOfAdvertisers}$ for all $i\in[\numberOfAdvertisers]$, return $\kipa{v}$\;}
 Set $s \leftarrow \max \{i\in[\numberOfAdvertisers]: \effv_i>0\}$\;
 \While{$(s- \numberOfSlots)g(\effv_s)\geq \sum_{i=1}^s g(\effv_i)$}{
  $s\leftarrow s-1$\;}
  For $i>s$: set $\kipaalloc_i = 0$\;
  For $i\leq s$ set $\kipaalloc_i = 1 - (s-k)\frac{g(\effv_i)}{\sum_{t=1}^s g(\effv_t)}$\;
  \Return $\kipa{v}$
 \caption{\kipaname}\label{alg:k-unit-IPA}
 \label{kunit}
\end{algorithm}
\vspace{0.5 cm}

\begin{flushleft}
\textbf{\Cref{thm:seperable-ipa-feasible}.} Let $M$ be the $n\times k$ matrix output by arbitrary run of \Cref{alg:IPA-separable} on input $(v, \userCTRVector, \slotCTRVector, k, g)$. Let $\mathcal{M}$ be the set of matchings from the set of advertisers $([\numberOfAdvertisers])$ to the set of slots $([\numberOfSlots])$. There exists a probability distribution $\mathcal{P}$ over $\mathcal{M}$ such that for all advertisers $i$ and slots $j$, the probability that $i$ is assigned to slot $j$ is equal to the probability prescribed by the matrix $M$, i.e. $$\Pr_{m\sim \mathcal{P}}[ (i, j) \in m] = M_{i,j}.$$ 

\end{flushleft}

\begin{proof}
By appealing to the Birkhoff-von Neumann algorithm, it suffices to show that we can extend $M$ to a doubly stochastic matrix $n\times n$ matrix $A$ in which the first $n$ rows and $k$ columns of $A$ are the matrix $M$. We can interpret slots $\numberOfSlots+1$ to $\numberOfAdvertisers$ as dummy slots, and whenever an advertiser is assigned to them, we interpret that to mean the ad is not shown in any slot. Now, we define $A$ as follows:
\[   
A_{i,j} = 
     \begin{cases}
        M_{i,j} &\text{ if } j \leq k \\     \frac{1-\kipah{k}_i}{\numberOfAdvertisers - \numberOfSlots } &\text{ if } j>k
     \end{cases}
\]

Now, we must show that $A$ is doubly stochastic. We show that each entry is in $[0,1]$ and then that each column and row sum is $1$.

First, observe that each entry of $A$ is in $[0,1]$. Each entry of $M$ is non-negative since $\kipah{j}_i \geq \kipah{j-1}_i$ because \Cref{alg:k-unit-IPA} is non-decreasing in $k$. These entries are also at most $1$ since the $\kipah{h}_i$ values are at most $1$. Finally, the entries of $A$ for $j > k$ (if such columns exist) are in the range $[0, 1]$ as well since they are fractions with numerators in $[0,1]$ and denominators at least $1$.

In addition, notice that for $j\leq k$, the $j$th column $\kipah{j} - \kipah{j-1}$ sums to one since $\kipah{h}$ sums to $h$. For $j >k$, $\sum_{i=1}^n (1 - \kipah{h}_i) / (n-k) = 1$ since the sum of the $\kipah{k}_i$ equals $k$. Finally, in each row $i$, the sum of the first $k$ columns is $\kipah{k}_i$, while the sum of the last $n-k$ columns is $1 - \kipah{k}_i$, so each row sums to $1$. 
\end{proof}\label{feasibility proof}


\subsection{Fairness}
\begin{flushleft}
\textbf{\Cref{lemma:kipa-value-stable}.}
    For the \kipaname\ mechanism with parameter $\ell$ run on any $k$ and any bid vectors $v$ and $v'$ with $\lambda = \max_{i \in [n]} \{\hat{v}_i / \hat{v}_i' , \hat{v}_i'/ \hat{v}_i\}$, for all indices $i$, $|\kipaalloc_i(v) - \kipaalloc_i(v') | \leq f_\ell(\lambda)$.
\end{flushleft}
\begin{proof}
The proof of this lemma is a generalization of and very similar to the proofs in \cite{CJ22} for \textit{Theorem 11} and \textit{Lemma 9, Lemma 14} and \textit{Lemma 15}. The proofs of the lemmas used in this proof can be found after the proof of lemma. The proof is as follows:

Let  $\lambda\in[1,\infty)$. We fix the value vector $v$, and find a vector $v'$ for which the difference $|a_1-a'_1|$ is maximized among all choices of $v'$ such that $\lambda$ is the maximum multiplicative difference between all indices. Note that in this proof we do not have the assumption that $v_1\geq\ldots\geq v_n$, so showing $|a_1-a'_1|\leq f_{\ell}(\lambda)$ concludes the proof.

The first step is to find the described $v'$ vector of values. Consider the following lemma:
\begin{lemma}\label{monotonicity}
    For any value vector $v$ and any $i\in[n]$, let $v'=(v'_i,v_{-i})$ be another value vector that differs from $v$ only in coordinate $i$ with $v_i>v'_i$. Then it holds that $a_i(v)\geq a_i(v')$ and $a_j(v)\leq a_j(v')$ for all $j\neq i$.
\end{lemma}
Using the monotonicity property implied by this lemma, we understand that the vector $v'$ we where looking for is defined as $v'_{-1}=v_{-1}$, $v'_1=\frac{v_1}{\lambda^2}$. In other words, $v'$ is the same as $v$ except in the first coordinate where it has the maximum possible difference. Now using \Cref{monotonicity}, we know that $a_1\geq a'_1$, and $a_j\leq a'_j$ for all $j\neq 1$. Let $S$ be the set of advertisers who receive non-zero allocation in $a$, and $S'$ be the corresponding set for $a'$. We know that $S\subseteq S'\cup\{1\}$. We want to show $a_1-a'_1 \leq f_{\ell}(\lambda)$. We consider three cases:

\textbf{Case 1: $1\notin S$. }In this case $a_1=0$ and since $a_1\geq a'_1$, then $a_1=a'_1=0$ and so $a_1-a'_1=0\leq f_{\ell}(\lambda)$.

\textbf{Case 2: $1\in S, 1\notin S'.$ }Consider the following lemmas:
\begin{lemma}\label{increase_advertisers}
    Let $v$ be any value vector with the advertisers reordered so that $v_1\geq\ldots\geq v_n$. Let $S=[m]$ be the set of advertisers with non-zero allocation returned by the $k-$unit IPA mechanism parametrized by function $g$ on $v$. Then for any $m'$ with $m\leq m'\leq n$ and any $i\in[m]$,
    $$1-(m-k)\frac{g(v_i)}{\sum\limits_{j\in [m]} g(v_j)}\leq 1-(m'-k)\frac{g(v_i)}{\sum\limits_{j\in [m']} g(v_j)}$$
\end{lemma}
\begin{lemma}\label{max-diff}
    Let $\lambda\in[1,\infty)$ and $g(x)=x^{-\ell}$ for all $x\geq 0$. Let $S\subseteq[n]$ be an arbitrary set of advertisers and $i$ be an adveriser in $S$. Suppose that value vectors $v$ and $v'$ satisfy $v'_i=\frac{v_i}{\lambda^2}$ and $v'_j=v_j$ for all $j\in S$ with $j\neq i$. Then we have:
   $$\max\left(0,1-(|S|-k)\frac{g(v_i)}{\sum\limits_{j\in S} g(v_j)}\right)-\max\left(0,1-(|S|-k \frac{g(v'_i)}{\sum\limits_{j\in S} g(v'_j)}\right)\leq \f$$
\end{lemma}
Using \Cref{increase_advertisers}, because $S\subseteq S'$, we get that
$$a_1=1-(|S|-k)\frac{g(v_i)}{\sum\limits_{j\in S} g(v_j)}\leq 1-(|S'|-k)\frac{g(v_i)}{\sum\limits_{j\in S'} g(v_j)}$$
Since $1\in S\cup S'$, then $a_1>0$. Therefore, $1-(|S'|-k)\frac{g(v_i)}{\sum\limits_{j\in S'} g(v_j)}$. Using this and \Cref{max-diff} for set $S'$, we get:
$$a_1\leq 1-(|S'|-k)\frac{g(v_i)}{\sum\limits_{j\in S'} g(v_j)}\leq \max(0,1-(|S'|-k)\frac{g(v'_1)}{\sum\limits_{j\in S'}g(v'_j)})+f_{\ell}(\lambda)=a'_1+f_{\ell}(\lambda)$$
and therefore $a_1-a'_1\leq f_{\ell}(\lambda)$.

\textbf{Case 3: $1\in S, 1\notin S'$. }We cannot follow the logic of case 2, because $1\notin S$. Instead, we define a new set $R$ and show that $S\subseteq R$. Let $R=\{i|v'_i\geq v'_1\}=\{i|v_i\geq \frac{v_1}{\lambda^2}\}$. Now note that since $1\notin S'$. So the value of all those advertisers who received non-zero allocation in $v'$ has to be at least $v'_1$. So for all $i\in S'$, $v'_i\geq v'_1$. Therefore, $S\subseteq S'\cup\{1\}\subseteq R$. Now using \Cref{increase_advertisers} on sets $S$ and $R$, or $|S|\leq |R|$ respectively, we get:
$$a_1=1-(|S|-k)\frac{g(v_1)}{\sum\limits_{j\in S}g(v_j)}\leq 1- (|R|-k)\frac{g(v_1)}{\sum\limits_{j\in R}g(v_j)}$$
Again, since $1\in S$, $a_1>0$ and so using \Cref{max-diff} we get:
$$1-(|R|-k)\frac{g(v_1)}{\sum\limits_{j\in R}g(v_j)}\leq \max(0, 1- (|R|-k)\frac{g(v'_1)}{\sum\limits_{j\in R}g(v'_j)})+\f $$
Here, we claim that $1-(|R|-k)\frac{g(v'_1)}{\sum\limits_{j\in R}g(v'_j)}\leq 0$. To show this, note that $S\subseteq R$, $S'\neq R$ and $v_1=\min(x|x\in R)$. So in the while loop in $k-$unit algorithm when determining which advertisers get allocation zero for $v'$, the set at question is at some point $R$, and since $1\notin S'$ and $1\in R$, it should hold that $(|R|-k)g(v'_1)\geq \sum\limits_{j\in R} g(v'_j)$, because $1$ is eliminated from $S'$. Therefore, $a'_1=0$. So using \Cref{increase_advertisers} and \Cref{max-diff} we get:
$$a_1\leq 1-(|R|-k)\frac{g(v_1)}{\sum\limits_{j\in R}g(v_j)}\leq a'_1+\f=\f$$
Therefore, $a_1-a'_1\leq \f$. This concludes the proof of lemma.
\end{proof}
In the rest of this subsection, we will see proofs of the lemmas used in proof of \Cref{lemma:kipa-value-stable}.

\begin{flushleft}
\textbf{\Cref{monotonicity}. }
    For any value vector $v$ and any $i\in[n]$, let $v'=(v'_i,v_{-i})$ be another value vector that differs from $v$ only in coordinate $i$ with $v_i>v'_i$. Then it holds that $a_i(v)\geq a_i(v')$ and $a_j(v)\leq a_j(v')$ for all $j\neq i$.
\end{flushleft}
\begin{proof}
One way of formulating the $k-$unit algorithm is as follows: let $y_i(t)=\max(0,1-tkg(v_i))$ and $y(t)=\sum\limits_{i=1}^n y_i(t)$. The allocations will be specified by finding a $t$ for which $y(t)=k$. Now define $y(t)$ and $y'(t)$ for $v$ and $v'$ respectively. Note that since $g$ is a decreasing function, we will have that $g(v_j)=g(v'_j)$ for all $j\neq i$, and $g(v_i)<g(v'_i)$. Therefore, for all $t\geq 0$ we have that $y(t)\geq y'(t)$. This means the critical $t$ that defines allocations in $k-$unit algorithm for $v$ should be less than for $v'$. Therefore, for every $j\neq i$, $a_j\leq a'_j$. Since both allocations sum to $k$ for the $n$ advertisers, it should also hold that $a_i\geq a'_i$. This concludes the lemma.
\end{proof}
\begin{flushleft}
\textbf{\Cref{increase_advertisers}. }
Let $v$ be any value vector with the advertisers reordered so that $v_1\geq\ldots\geq v_n$. Let $S=[m]$ be the set of advertisers with non-zero allocation returned by the $k-$unit IPA mechanism parametrized by function $g$ on $v$. Then for any $m'$ with $m\leq m'\leq n$ and any $i\in[m]$,
    $$1-(m-k)\frac{g(v_i)}{\sum\limits_{j\in [m]} g(v_j)}\leq 1-(m'-k)\frac{g(v_i)}{\sum\limits_{j\in [m']} g(v_j)}$$
\end{flushleft}
\begin{proof}
Let $a$ be the vector of allocations produced by $k-$unit algorithm on $v$, and define $a'_i=1-(m'-k)\frac{g(v_i)}{\sum\limits_{j\in[m']}g(v_j)}$ for all $i\in[m']$. If we use the same formulation as mentioned in proof of \Cref{monotonicity}, since for all $j\notin[m]$, $a_j=0$, then we get that $y_t(v)=\sum\limits_{i\in[m]} y_i(t)=\sum\limits_{i\in[m']}y_i(t)$. Let $t^*$ be the value for $t$ for which $y_t(v)=k$. Moreover, if we define $a'_i(t)=1-(m'-k)\frac{g(v_i)}{\sum\limits_{j\in [m']} g(v_j)}$, $y'_i(t)=1-tkg(v_i)$ and $y'(t)=\sum\limits_{i\in [m']}y'_i(t)$, the value of $t$ for which $y'(t)=k$ will be some $t'$ with $t'\leq t^*$. This is because values of $y'_i(t)$ can be also negative. So as we increase $t$ from $0$, we hit the $k$ mark earlier in $y'(t)$ than $y(t)$. Therefore, we will have that for all $i\in[m]$,
$$a_i=1-(m-k)\frac{g(v_i)}{\sum\limits_{j\in [m]} g(v_j)}=1-t^*kg(v_i)\leq 1-t'kg(v_i)= a'_i=1-(m'-k)\frac{g(v_i)}{\sum\limits_{j\in [m']} g(v_j)}$$
This concludes the proof of lemma.

\end{proof}
\begin{flushleft}
\textbf{\Cref{max-diff}. }Let $\lambda\in[1,\infty)$ and $g(x)=x^{-\ell}$ for all $x\geq 0$. Let $S\subseteq[n]$ be an arbitrary set of advertisers and $i$ be an adveriser in $S$. Suppose that value vectors $v$ and $v'$ satisfy $v'_i=\frac{v_i}{\lambda^2}$ and $v'_j=v_j$ for all $j\in S$ with $j\neq i$. Then we have:
    $$\max\left(0,1-(|S|-k)\frac{g(v_i)}{\sum\limits_{j\in S} g(v_j)}\right)-\max\left(0,1-(|S|-k \frac{g(v'_i)}{\sum\limits_{j\in S} g(v'_j)}\right)\leq \f$$
\end{flushleft}
\begin{proof}
This proof follows the proof of a similar lemma in \cite{CJ22}. We are interested to show
\begin{equation*}
    d=\max\left(0,1-(|S|-k)\frac{g(v_i)}{g(v_i)+\sum\limits_{j\in S, j\neq i} g(v_j)}\right)-\max\left(0,1-(|S|-k \frac{g(\frac{v_i}{\lambda^2})}{g(\frac{v_i}{\lambda^2})+ \sum\limits_{j\in S, j\neq i} g(v_j)}\right)\leq \f
\end{equation*}
If the maximum in the first term of $d$ is zero, or $|S|=k$, the inequality is trivially true. So we may assume that $|S|\geq k+1$ and $1-(|S|-k)\frac{g(v_i)}{g(v_i)+\sum\limits_{j\in S, j\neq i} g(v_j)}>0$. Let $r=\frac{\sum\limits_{j\in S, j\neq i} g(v_j)}{g(v_i)}$. Moreover, we know that $g(\frac{v_i}{\lambda^2}=\lambda^{2\ell}g(v_i)$. So
\begin{equation*}
    d(r)=1-\frac{|S|-k}{1+r}-\max(0,1-\frac{|S|-k}{1+\lambda^{-2\ell}r})=\frac{|S|-k}{\max(|S|-k, 1+\lambda^{-2\ell}r)}-\frac{|S|-k}{1+r}
\end{equation*}
Consider the following three cases:

\textbf{Case 1: }$|S|=k+1$. Then $d(r)=\frac{1}{1+\lambda^{-2\ell}r}-\frac{1}{1+r}$. This expression is maximized at $r=\lambda^\ell$ and so $d(r)\leq \frac{\lambda^\ell-1}{\lambda^\ell+1}\leq 1-\lambda^{-\ell}$.

\textbf{Case 2: }$|S|>k+1$ and $\max(|S|-k, 1+\lambda^{-2\ell}r=|S|-k$. So $r\leq \lambda^{2\ell}(|S|-k-1).$ Then $d(r)=1-\frac{|S|-k}{1+r}$ and is an increasing function of $r$. So it is maximized at $r=\lambda^{2\ell}(|S|-k-1)$ for which
\begin{equation*}
    d(r)\leq 1- \frac{|S|-k}{1+\lambda^{2\ell}(|S|-k-1)}\leq 1-\lambda^{-2\ell}=\f
\end{equation*}

\textbf{Case 3: }$|S|>k+1$ and $\max(|S|-k,1+\lambda^{-2\ell}r)=1+\lambda^{-2\ell}r$.  Then
\begin{equation*}
    d(r)=(|S|-k)(\frac{1}{1+\lambda^{-2\ell}r}-\frac{1}{1+r}
\end{equation*}
Now $d(r)$ is non-increasing on $r\in[\lambda^{2\ell}(|S|-k-1), \infty)$ and so
\begin{equation*}
    d(r)\leq 1- \frac{|S|-k}{1+\lambda^{2\ell}(|S|-k-1)}\leq 1-\lambda^{-2\ell}=\f
\end{equation*}
This concludes proof of the lemma.
\end{proof}
\begin{flushleft}

\subsection{Social Welfare}
\textbf{\Cref{thm:sep-IPA-value-stable}. }
   The \mipaname\ mechanism with parameter $\ell > 0$ and for any number of advertisers $\numberOfAdvertisers$ is \weakvalueStable\ with respect to any function $f$ satisfying $f(\lambda) \geq f_\ell(\lambda) = 1 - \lambda^{-2\ell}$ for all $\lambda \in [1, \infty)$, as in \Cref{def:weakly-value-stable}.
\end{flushleft}
\begin{proof}
Fix the parameter $\ell$ and consider some $i\in [\numberOfAdvertisers]$ and  $j \in [\numberOfSlots]$. By \Cref{lemma:kipa-value-stable}, for all $h \in [k]$, $\abs{\kipah{h}_i(v) - \kipah{h}_i(v')} \leq f_\ell(\lambda)$.

\begin{align*}
    |M_{i,j}(v) - M_{i,j}(v')| &= \abs{ \left[ \kipah{j}_i(v) - \kipah{j-1}_i(v) \right] - \left[ \kipah{j}_i(v') - \kipah{j-1}_i(v') \right] } \\
    &= \abs{ \left[ \kipah{j}_i(v) - \kipah{j}_i(v') \right] + \left[ \kipah{j-1}_i(v') - \kipah{j-1}_i(v) \right] } \\
    &\leq \abs{ \kipah{j}_i(v) - \kipah{j}_i(v') } + \abs{ \kipah{j-1}_i(v') - \kipah{j-1}_i(v) } \\
    &\leq f_\ell(\lambda) +  f_\ell(\lambda) \\
    &= 2f_\ell(\lambda)
\end{align*}
\end{proof}

 \begin{flushleft}
 \textbf{\Cref{thm:ipa-ordered-value-stability}. } 
     \mipaname\ with parameter $\ell$ satisfies \hfairness\ with respect to $f_\ell(\lambda)$. That is, for every set of value and CTR vectors $v$, $v'$, $\userCTRVector$, $\userCTRVector'$ and $\slotCTRVector$, as well as for any advertiser $i$ and any decreasing vector $h$ with 
     $1 \geq h_{1} \geq \ldots \geq h_{k} \geq 0$:
 
$$\abs{ \sum_{j =1}^k h_{j} \left(\mipa_{i,j}-\mipa'_{i,j}\right) } \leq f_{\ell}(\lambda) \text{ where } \lambda \text{ is defined as } \max_{i\in[n]} \left( \max\left\{\frac{\alpha_iv_i}{\alpha'_iv'_i}, \frac{\alpha'_iv'_i}{\alpha_iv_i}\right\}\right)$$
where
$\mipa= \separableMechanism{v,\userCTRVector,\slotCTRVector}$ and $\mipa'= \separableMechanism{v',\userCTRVector',\slotCTRVector}$.
 \end{flushleft}

\begin{proof}
    Fix some vectors $v, v', \alpha, \alpha',$ and $\beta$, and the corresponding allocation matrices $M$ and $M'$. Consider some advertiser $i$. We begin by using the definition \mipaname\ and then rearranging terms. Note that we define $h_{k+1} \coloneqq 0$ for notational simplicity. 
    \begin{align*}
        \abs{ \sum_{j =1}^k h_{j} \left(\mipa_{i,j}-\mipa'_{i,j}\right) } &= \abs{ \sum_{j =1}^k h_{j} \left((\kipah{j}_i-\kipah{j-1}_i) - (\kipah{j}_{i'}-\kipah{j-1}_{i'}) \right) } \\
        &= \abs{ \sum_{j =1}^k \left( h_{j} (\kipah{j}_i-\kipah{j-1}_i) - h_{j} (\kipah{j}_{i'}-\kipah{j-1}_{i'}) \right) } \\
        &= \abs{ \sum_{j =1}^k \left(\kipah{j}_i-\kipah{j-1}_{i'}\right) \left(h_{j} - h_{j+1} \right) }
    \end{align*}
    Now, observe that because $h_1 \leq 1$ and the coefficients $(h_{j} - h_{j+1})$  telescope, the sum of these coefficients is at most $1$. Since the expression is a weighted sum over columns of the differences in allocation at that column, the expression is bounded by the maximum difference in any column. But because \mipaname\ satisfies value stability (by \Cref{lemma:kipa-value-stable}), this is bounded by $f_\ell(\lambda)$, as desired. 
    
    $$\abs{ \sum_{j =1}^k h_{j} \left(\mipa_{i,j}-\mipa'_{i,j}\right) } = \abs{ \sum_{j =1}^k \left(\kipah{j}_i-\kipah{j-1}_{i'}\right) \left(h_{j} - h_{j+1} \right) } = \abs{ \max_j \left( \kipah{j}_i-\kipah{j-1}_{i'} \right) } \leq f_\ell(\lambda)$$
\end{proof}

\begin{flushleft}
\textbf{\Cref{thm:ipa-separable-sw}. }
    The IPA algorithm for the separable case, \Cref{alg:IPA-separable}, run with parameter $\ell > 0$ and any number of advertisers $\numberOfAdvertisers$ achieves a $\left(1-\frac{\ell^{\ell}}{(1+\ell)^{\ell+1}}\right)$ approximation the social welfare of the unfair optimum.
\end{flushleft}

\begin{proof}
Suppose the \kipaname\ mechanism attains an $\eta$ approximation to the optimal social welfare in the \kunitsetting. Then the \sipaname\ mechanism attains the same approximation factor $\eta$ in the \separableModel when run with the \kipaname mechanism as a subroutine. In order to prove this, we consider the social welfare attained by the \sipaname\ mechanism. 

    \begin{align*}
     \swalg &=\sum_{i =1}^n \sum_{j=1}^k \userCTR_i v_i \slotCTR_j M_{ij} \\ 
     &=\sum_{i=1}^n \sum_{j=1}^k \effv_i \slotCTR_j \left[\kipah{j}_i - \kipah{j-1}_i\right]\\ 
     &=\sum_{i=1}^n \sum_{j=1}^k \effv_i(\slotCTR_j-\slotCTR_{j+1}) \kipah{j}_i \text{since } \slotCTR_{k+1} = 0 \text{ and } \kipah{0}_i= \zerovector 
\end{align*}

\noindent
Since for all $j\in [k]$, $\sum_{i} \effv_ia^{(j)}_i \ge \eta (\effv_1 + \cdots + \effv_j)$, then:
\begin{align*}
    \swalg 
    &= \sum_{j=1}^k(\slotCTR_j-\slotCTR_{j+1}) \left(\sum_{i=1}^n \effv_i \kipah{j}_i \right)\\
    &\geq \eta \sum_{j=1}^k (\slotCTR_j-\slotCTR_{j+1})\left(\effv_1 + \cdots + \effv_j\right) \\ 
    &= \eta \sum_{j=1}^k \effv_j\slotCTR_j \\
    &= \eta \, \swopt
\end{align*}
Finally, we know by \Cref{lemma:k-unit-IPA-SW} that the \kipaname\ mechanism is an $\eta=\left(1-\frac{\ell^{\ell}}{(1+\ell)^{\ell+1}}\right)$-approx -imation to the optimal $k$-unit social welfare. 
Replacing $\eta$ by $\left(1-\frac{\ell^{\ell}}{(1+\ell)^{\ell+1}}\right)$ concludes the proof.
\end{proof}

\begin{flushleft}
\textbf{\Cref{ipa-sw-l=1}.} The IPA algorithm for the $k-$unit case, \Cref{alg:k-unit-IPA}, run with parameter $\ell=1$ and any number of advertisers $\numberOfAdvertisers$ achieves a $\frac{3}{4}$-approximation to the social welfare of the unfair optimum.
\end{flushleft}
\begin{proof}
Let $s\leq n$ be the number of advertisers who receive a non-zero allocation, and denote their values by $\effv_1\geq \ldots \geq \effv_s$. We will have:
$$a_i=1-(s-k)\frac{\effv_i^{-1}}{\sum_{j=1}^s \effv^{-1}_j}$$ 
Now for any $i\geq 1$ we can write:
$$1> 1-a_s=(s-k)\frac{\effv_s^{-1}}{\sum_{j=1}^s \effv_j^{-1}} \Rightarrow \sum_{j=1}^s \effv_j^{-1}>\frac{s-k}{\effv_s}$$
Using this, for all $i\geq 1$ we can write:
$$a_i=1-(s-k)\frac{\effv_i^{-1}}{\sum_{j=1}^s \effv^{-1}_j}> 1-(s-k)\frac{v_s}{(s-k)\effv_i}=1-\frac{\effv_s}{\effv_i}$$
Now we will examine the social welfare of the algorithm:
$$\swalg=\sum_{i=1}^s a_i \effv_i > \sum_{i=1}^k (1-\frac{\effv_s}{\effv_i})\effv_i+(\sum_{i=1}^k \frac{1}{\effv_i})\effv_s^2=\sum_{i=1}^k \effv_i-k\effv_s+\effv_s^2(\sum_{i=1}^k \frac{1}{\effv_i})$$
Now using Cauchy-Shwarz inequality, we have: $(\effv_1+\ldots+\effv_k)(\frac{1}{\effv_1}+\ldots+\frac{1}{\effv_k})\geq k^2$. Moreover, note that $\swopt=\sum_{i=1}^k \effv_i$. Let $t=\effv_1+\ldots+ \effv_k$. So we will have:
$$\frac{\swalg}{\swopt}> \frac{\sum_{i=1}^k \effv_i + \effv_s^2(\sum_{i=1}^k \frac{1}{\effv_i})-k\effv_s}{\sum_{i=1}^k \effv_i} = 1+\frac{\effv_s^2(\frac{k^2}{t})-k\effv_s}{t}=1+\frac{\effv_s k}{t}^2-\frac{k\effv_s}{t}$$
Now we can minimize the ratio. Note that since for $1\leq i\leq k, \effv_i\geq \effv_s$, we should have $t\geq k\effv_s$. 
$$\frac{\swalg}{\swopt} > \min \left[ \left(\frac{k\effv_s}{t}-\frac{1}{2}\right)^2+\frac{3}{4} \right] \quad \text{ s.t. } \effv_s\geq 0, t\geq k\effv_s$$
We can see that the ratio has the minimum of $\frac{3}{4}$, which is achieved at $v_s=\frac{t}{2k}$ so that the first term is zero.
\end{proof}

\begin{flushleft}
\textbf{\Cref{lemma:k-unit-IPA-SW}. }
    The IPA algorithm for the $k-$unit case, \Cref{alg:k-unit-IPA}, run with parameter $\ell$ and any number of advertisers $\numberOfAdvertisers$ achieves a $\left(1-\frac{\ell^{\ell}}{(1+\ell)^{\ell+1}}\right)$-approximation to the social welfare of the unfair optimum. 
\end{flushleft}
\begin{proof}
 Let $m$ be the number of agents allocated a non-zero value by the $k-$unit algorithm and $\hat{v}_1\geq \ldots \geq\hat{v}_n$. We would like to find a lower bound on $\frac{\sum_{i=1}^m \kipaalloc_i\hat{v}_i}{\sum_{i=1}^k \hat{v}_i}$. Note that we can write:
$$\sum_{i=1}^m \kipaalloc_i\hat{v}_i = \sum_{i=1}^k \kipaalloc_i\hat{v}_i + \sum_{i=k+1}^m \kipaalloc_i\hat{v}_i \geq \sum_{i=1}^k \kipaalloc_i\hat{v}_i + \hat{v}_m\left(1-\sum_{i=1}^k \kipaalloc_i\right)$$
This is because for all $i\leq m, a_i\geq a_m$. Moreover, we have:
\begin{equation}\label{vm}
    1>1-\kipaalloc_m = (m-k)\frac{\hat{v}_m^{-\ell}}{\sum_{t=1}^m \hat{v}_t^{-\ell}} \Rightarrow a_i\geq 1-\left(\frac{\hat{v}_m}{\hat{v}_i}\right)^{\ell}
\end{equation}

\remove{Using inequality \ref{vm} one can write:
\begin{equation}
    R\geq \frac{\sum_{i=1}^k \hat{v}_i(1-(\frac{\hat{v}_m}{\hat{v}_i})^{\ell})+\hat{v}_m(\sum_{i=1}^k (\frac{\hat{v}_m}{\hat{v}_i})^{\ell})}{\sum_{i=1}^k \hat{v}_i} = \frac{\sum_{i=1}^k \hat{v}_i - \hat{v}_m^{\ell}(\sum_{i=1}^k \hat{v}_i^{-\ell+1})+\hat{v}_m^{\ell+1}(\sum_{i=1}^k \hat{v}_i^{-\ell})}{\sum_{i=1}^k \hat{v}_i}
\end{equation}}

Note that we can assume $\sum\limits_{i=1}^k \hat{v}_i=1$ because the allocation is scale-free. Using this and \Cref{vm} we can write:
$$\swalg= \swopt-\hat{v}_m^\ell\left(\sum_{i=1}^k \hat{v}_i^{-\ell+1}\right)+\hat{v}_m^{\ell+1}\left(\sum_{i=1}^k \hat{v}_i^{-\ell}\right)$$

We show the theorem for the case of $\ell=1$ separately in the following lemma the proof of which is deferred to \Cref{sec:appendix-ipa}. 
\begin{lemma}\label{ipa-sw-l=1}
     The IPA algorithm for the $k-$unit case, \Cref{alg:k-unit-IPA}, run with parameter $\ell=1$ and any number of advertisers $\numberOfAdvertisers$ achieves a $\frac{3}{4}$-approximation to the social welfare of the unfair optimum.
\end{lemma}

So from now on we may assume $\ell\neq 1$. Now we are interested in finding a lower bound for ratio of \swopt\ to \swalg . Note that \swopt\ is independent of $\hat{v}_m$, so to minimize the ratio $\frac{\swalg}{\swopt}$, we can minimize $\swalg$ as a function of $\hat{v}_m$. Therefore, we now consider the derivative of $\swalg$ with respect to $\hat{v}_m$. 

\begin{equation}
    \frac{\partial \swalg}{\partial \hat{v}_m}= -\ell \left(\sum_{i=1}^k \hat{v}_i^{-\ell+1}\right) \hat{v}_m^{\ell-1}+(\ell+1)\left(\sum_{i=1}^k \hat{v}_i^{-\ell}\right)\hat{v}_m^{\ell}=0
\end{equation}

Since we can assume that $\hat{v}_m > 0$, because the algorithm allocates $0$ to any advertiser with effective value of $0$, we can divide the above equation by $(\hat{v}_m)^{\ell-l}$ and solve for $\hat{v}_m$.
\begin{equation}
    \hat{v}_m= \frac{\sum_{i=1}^k \hat{v}_i^{-\ell+1}}{\sum_{i=1}^k \hat{v}_i^{-\ell}}\times \frac{\ell}{\ell+1}
\end{equation}
Hence, the $\frac{\partial \swalg}{\partial \hat{v}}$ is $0$ at the above value. The second derivative is strictly positive here, so this is indeed the minimum in the region $\hat{v}_m>0$. 

Now we plug this value for $\hat{v}_m$ into our expression for \swalg, getting:
\begin{equation}
    \swalg=\swopt \left(1- \frac{(\sum_{i=1}^k \hat{v}_i^{-\ell+1})^{\ell+1}}{(\sum_{i=1}^k \hat{v}_i^{-\ell})^{\ell}(\sum_{i=1}^k \hat{v}_i)}\times \frac{\ell^{\ell}}{(\ell+1)^{\ell+1}}\right)
\end{equation}
Now, since the term  $\frac{\ell^{\ell}}{(\ell+1)^{\ell+1}}$ depends only on $\ell$ and not the $\hat{v}_i$, it suffices to maximize the term $\frac{(\sum_{i=1}^k \hat{v}_i^{-\ell+1})^{\ell+1}}{(\sum_{i=1}^k \hat{v}_i^{-\ell})^{\ell}(\sum_{i=1}^k \hat{v}_i)}$ as a function of the $\hat{v}_i$:
$$\frac{(\sum_{i=1}^k \hat{v}_i^{-\ell+1})^{\ell+1}}{(\sum_{i=1}^k \hat{v}_i^{-\ell})^{\ell}(\sum_{i=1}^k \hat{v}_i)}$$
Next, we use the following lemma with its proof deferred to appendix.
\begin{lemma}\label{lemma:sw-ipa-maximize}
    The function 
    $$\frac{(\sum_{i=1}^k \hat{v}_i^{-\ell+1})^{\ell+1}}{(\sum_{i=1}^k \hat{v}_i^{-\ell})^{\ell}(\sum_{i=1}^k \hat{v}_i)}$$
    for any $\ell>0, \ell\neq 1$ and $\hat{v_1}\geq \ldots\geq \hat{v_k}$ is maximized when $\hat{v_1}=\ldots=\hat{v_k}$
\end{lemma}
Now using this lemma, at the maximized point we have $\hat{v}_1=\ldots=\hat{v}_k$. So we get:
$$\swalg=\swopt\left(1-\frac{k^{\ell+1}}{k^{\ell}(k)}\times \frac{\ell^\ell}{(\ell+1)^{\ell+1}}\right)=\swopt(1-\frac{\ell^\ell}{(\ell+1)^{\ell+1}})$$
This concludes the proof of this theorem.
\end{proof}

\begin{flushleft}
\textbf{\Cref{lemma:sw-ipa-maximize}.}
  The function 
    $$\frac{(\sum_{i=1}^k \hat{v}_i^{-\ell+1})^{\ell+1}}{(\sum_{i=1}^k \hat{v}_i^{-\ell})^{\ell}(\sum_{i=1}^k \hat{v}_i)}$$
    for any $\ell>0, \ell\neq 1$ and $\hat{v_1}\geq \ldots\geq \hat{v_k}$ is maximized when $\hat{v_1}=\ldots=\hat{v_k}$.
\end{flushleft}
\begin{proof}
First, for all $1\leq i\leq k$, we define $\hat{v}_i=c_i\hat{v}_k$. Because $\hat{v}_1\geq \ldots \geq \hat{v}_k$, we know that $c_1\geq \ldots \geq c_{k-1}\geq c_k= 1$. Moreover, define $\csum{\ell} = \sum_{i=1}^{k} c_i^{-\ell} $. Now what we want to maximize can be written as:
\begin{equation}
    \text{maximize } f_{\ell}(c_1,\ldots, c_{k}) = \frac{\csum{\ell-1}^{\ell+1}}{\csum{\ell}^{\ell}\csum{-1}}
\end{equation}
Now computing the derivative, we get:
\begin{equation}
    g(\ell, i) = \frac{\partial f_{\ell}(c_1,\ldots, c_{k})}{\partial c_i} = \frac{(\csum{\ell})^{\ell-1}(\csum{\ell-1})^{\ell} A_i}{(\csum{\ell})^{2\ell}(\csum{-1})}
\end{equation}
where
\begin{equation}
    A_i=\csum{\ell}\csum{-1}(1-\ell^2)c_i^{-\ell} - \csum{\ell-1} \left[\csum{\ell}-\ell^2c_i^{-\ell-1}\csum{-1}\right]
\end{equation}

We are interested in the maximum feasible value of $ f_{\ell}(c_1,\ldots, c_{k})$. For each $1\leq i\leq k$, value of $g(\ell,i)$ can be positive, negative or zero. But note that since we are looking for the maxima of $f$, if for some $i$ we have $g(\ell,i)>0$, then in a maxima point we should have $c_i=c_{i-1}$. Equivalently, if $g(\ell,i)<0$, we must have that $c_i=c_{i+1}$. Finally, the sign of $A_i$ and $g(\ell,i)$ is the same. Similarly, since the other factors are identical for all $i$, $\frac{f_{\ell}(c_1,\ldots, c_{k-1})}{\partial c_i} = \frac{f_{\ell}(c_1,\ldots, c_{k-1})}{\partial c_j}$ if and only if $A_i = A_j$. In addition, observe that if $c_i = c_j$, then $A_i = A_j$. 

We claim that at a maxima point, there are some $i \leq j $ such that we have $c_1 = \dots = c_i$ with $A_1 > 0$, $c_{i+1} =\dots =c_j$ with $A_{j} = 0$, and $c_{j+1}= \dots =c_{k}= 1$ with $A_{j+1} < 0$. Note that we permit degenerate cases where $i = 0$ or $i = j$. So essentially, the worst-case effective value bid vector for \swalg is in the form $(a, \dots, a, b, \dots, b, 1, \dots, 1)$, where the partials with respect to the $a$ bids are positive, with respect to the $b$ bids are $0$, and with respect to the $1$ bids are negative. 

This is true because if the derivative with respect to some $c_i$ is negative, $c_i$ can be lowered until either the derivative hits $0$ or $c_i$ hits its lower bound, which is $c_{i+1}$. Similarly, if the derivative with respect to some $c_i$ is positive, $c_i$ can be raised until either the derivative hits $0$ or $c_i$ hits its upper bound, which is $c_{i-1}$. Finally, we will show that if the derivatives with respect to two $c_i$ and $c_j$ are both $0$, then $c_i = c_j$.  

Assume that for some $i<j$, $g(\ell,i)=g(\ell,j)=0$.  Considering, equivalently, that $A_i=0$ and $A_{j}=0$ and dividing the corresponding equalities, we get:
\begin{equation}
    \frac{c_i^{-\ell}}{c_{j}^{-\ell}} = \frac{\csum{\ell} - \ell^2c_i^{-\ell-1}\csum{-1}}{\csum{\ell} - \ell^2c_{j}^{-\ell-1}\csum{-1}}
\end{equation}
Therefore,
\begin{equation}
    \csum{\ell}\left(c_i^{-\ell}-c_{j}^{-\ell}\right) = \frac{\ell^2}{c_i^{\ell+1}c_{j}^{\ell+1}}\left(c_i-c_{j}\right)\csum{-1}
\end{equation}
Now if $c_i>c_{j}$, then $c_i^{-\ell}<c_{j}^{-\ell}$. So LHS will be negative and RHS will be positive. Therefore, $c_i=c_{j}$. This means we also should have $c_i=\ldots = c_{j}.$ Next, consider the following lemma:
\begin{lemma}\label{ipa-sw-case analysis}
    Considering maxima points of $f_{\ell}(c_1,\ldots,c_k)$ and the partial derivatives $g(\ell,i)$ for all $1\leq i\leq k$, the only possible case is when for all $1\leq i\leq k$, $g(\ell,i)=0$.
\end{lemma}
Using this, since $g(\ell,i)=0$ for all $i$, this means $c_1=\ldots=c_k=1$, and so it has to be that $\hat{v_1}=\ldots=\hat{v_k}$ at the point of maximum. This concludes the proof of lemma.
\end{proof}

\begin{flushleft}
\textbf{\Cref{ipa-sw-case analysis}.}
    Considering maxima points of $f_{\ell}(c_1,\ldots,c_k)$ and the partial derivatives $g(\ell,i)$ for all $1\leq i\leq k$, the only possible case is when for all $1\leq i\leq k$, $g(\ell,i)=0$.
\end{flushleft}
\begin{proof}
The proof of this lemma is by considering corner cases and showing that we get a contradiction if they happen. These cases are as follows:
\begin{enumerate}
\item For some $i\geq 1$, $A_i>0$ and $A_{i+1}=0$.
\item For some $i\leq k$, $A_{i-1}=0$ and $A_i=0$.
\item For some $i\geq 1$, $A_i>0$ and $A_{i+1}<0$.
\item For all $i\in[k]$, $A_i>0$.
\item For all $i\in[k]$, $A_i<0$.
\end{enumerate}
We show cases 1 as an example. The rest have a similar proof.

\textbf{Case 1. }Let $\ell<1$. If for some $i\geq 1$, $A_i>0$ and $A_{i+1}=0$, we get the two following inequalities:
$$C_{\ell}C_{-1}(1-\ell^2)c_i^{-\ell}>C_{\ell-1}[C_\ell-\ell^2c_i^{-\ell-1}C_{-1}]$$
$$C_{\ell}C_{-1}(1-\ell^2)c_{i+1}^{-\ell}=C_{\ell-1}[C_\ell-\ell^2c_{i+1}^{-\ell-1}C_{-1}]$$
The two sides of the second equality are positive. Dividing the inequality by the two sides of the equality, we get:
$$(\frac{c_{i+1}}{c_i})^\ell>\frac{C_\ell-\ell^2c_i^{-\ell-1}C_{-1}}{C_\ell-\ell^2c_{i+1}^{-\ell-1}C_{-1}}$$
Note that since $c_i> c_{i+1}$, the LHS of the above inequality is less than $1$ while the RHS is more than $1$. This is a contradiction showing such case cannot happen. The same kind of argument goes for $\ell>1$.

Following the same logic, other cases can be shown. This concludes the proof.
\end{proof}

\section{Deferred Proofs From \Cref{sec:ef}}\label{sec:appendix-ef-proofs}
\begin{flushleft}
\textbf{\Cref{thm:ipa_weak_piif}}
    The \mipaname\ and \mpaname\ mechanisms $\separableMechanism{\cdot}$ with parameter $\ell$ are value-stable for similar users with heterogeneous preferences. 
\end{flushleft}

\begin{proof}
    Fix users with \alphasname\ $\userCTRVector$ and $\userCTRVector'$ and value vectors $v$ and $v'$. Also fix \betaname\ $\slotCTRVector$, advertiser $i$, and column $j$. Let $\mipa =\separableMechanism{v,\userCTRVector, \slotCTRVector}$ and $\mipa' =\separableMechanism{v,\userCTRVector', \slotCTRVector}$, where $\kipaalloc_t = \sum_{s=1}^{j} \mipa_{t,s}, \text{ and } \kipaalloc'_t = \sum_{s=1}^{j} \mipa'_{t,s}$. Finally, fix a permutation $\pi$ on advertisers for which $\frac{\userCTR_{\pi_1}}{\userCTR'_{\pi_1}}\geq \ldots \geq \frac{\userCTR_{\pi_{\numberOfAdvertisers}}}{\userCTR'_{\pi_{\numberOfAdvertisers}}}$.
    
    Since $\separableMechanism{\cdot}$ is envy-free, we know that $$\sum_{s = 1}^i \sum_{t=1}^j M_{st}(\alpha v) \geq \sum_{s = 1}^i \sum_{t=1}^j M_{st}(\alpha' v).$$ 
    
    Therefore, it suffices to show $$\sum_{s = 1}^i \sum_{t=1}^j M_{st}(\alpha' v) \geq \sum_{s = 1}^i \sum_{t=1}^j M_{st}(\alpha' v') - i f(\lambda)$$
    
    Consider the difference $$\sum_{s = 1}^i \sum_{t=1}^j M_{st}(\alpha' v') - \sum_{s = 1}^i \sum_{t=1}^j M_{st}(\alpha' v)$$
    
    Since $\sum_{t=1}^j M_{st}(\alpha v) = a^j_s(\alpha v)$, we can simplify this to:
    
    \begin{align*}
        \sum_{s = 1}^i a^j_s(\alpha' v') - \sum_{s = 1}^i a^j_s(\alpha' v) &\leq \abs{\sum_{s = 1}^i a^j_s(\alpha' v') - \sum_{s = 1}^i a^j_s(\alpha' v)}\\
        &= \abs{\sum_{s = 1}^i a^j_s(\alpha' v') - a^j_s(\alpha' v)}\\
        &\leq \sum_{s = 1}^i \abs{ a^j_s(\alpha' v') - a^j_s(\alpha' v)} \\
        &\leq \sum_{s = 1}^i f(\lambda) \\
        &= i * f(\lambda)
    \end{align*}
    
    Simply combining this with the previous inequality gives the desired result.
\end{proof}

\section{Deferred Proofs from \Cref{sec:pa}}\label{sec:appendix-pa}
\subsection{Algorithm}
Below is the algorithmic description of \separableModel in the case of $\Vec{\beta}=1$:

\vspace{0.5 cm}
\begin{algorithm}[H]
 \textbf{Input: }Vector $v$ of non-negative advertiser bids for user $u$; ad-specific CTRs $\userCTR_1, \cdots, \userCTR_n$; number of slots $\numberOfSlots$; function $g:\mathbb{R}^{\geq 0} \rightarrow [0,\infty]$ with $g$ a continuous, super-additive, increasing function and $g(0)=0$;\\
 \textbf{Initialization:}
 Determine effective values, WLOG assume $\effv_1 \geq \ldots\geq \effv_{\numberOfAdvertisers}$\;
 \If{$k=0$}
 {\Return $\kpa(v)=\overrightarrow{0}$}
 \If{$\effv_1\leq 0$}{
 Set $\kpaalloc_i = \frac{\numberOfSlots}{\numberOfAdvertisers}$ for all $i\in[\numberOfAdvertisers]$, return $\kpa(v)$\;}
 Set $s \leftarrow \max \{i\in[\numberOfAdvertisers]: \effv_i>0\}$\;
 Set $r=1$\;
 \While{$\frac{k.g(\effv_{r})}{\sum\limits_{t=r}^s g(\effv_{t})}\geq 1$}{$\kpaalloc_r=1$\;
 $r\leftarrow r+1$\;}
 For $i\geq r$: set $\kpaalloc_i = \frac{(k-r).g(\effv_i)}{\sum\limits_{t=r}^s g(\effv_{t})}$\;

  \Return $\kpa(v)$\;
 \caption{\kpaname}\label{PA-algorithm}
\end{algorithm}
\vspace{0.5 cm}

\subsection{Fairness}

\begin{flushleft}
\textbf{\Cref{lemma:k-unit-PA-fairness}.}
    The \kpaname\ mechanism with parameter $g(x) = x ^\ell$ satisfies total variation value stability with respect to $\ell$. 
    That is, for all pairs of effective value vectors $\effv, \effv'$ and subsets of advertisers $S \subseteq [n]$, 
    $$\abs{\sum_{s \in S} a_s(\effv) - \sum_{s \in S} a_s(\effv')}  \leq \frac{\lambda^\ell - 1}{\lambda^\ell + 1} \leq f_\ell(\lambda).$$
\end{flushleft}

\begin{proof}
    Fix some pairs of effective value vectors $\effv, \effv'$ and a subset of advertisers $S \subseteq [n]$. 
    Define $E$ to be $\sum_{s \in S} a_s(\effv) - \sum_{s \in S} a_s(\effv')$ and assume without loss of generality that $E \geq 0$. We want to upper bound $E$ by $f_\ell(\lambda)$.
    
    
    First, we reduce the general case to that where the while loop never executes. That is, we modify the given instance so that the while loop never executes while only increasing $E$ and decreasing $\lambda$. 
    First, we can assume that $i \in S$ if $a_i(\effv) > a_i(\effv')$ and $i \not\in S$ if $a_i(\effv) < a_i(\effv')$, since that those choices maximize $E$ (and do not effect $\lambda$). 
    We also assume that for all $i$, $\effv_i \geq \effv_i'$ and therefore $\lambda = \max_i \{\effv_i / \effv_i'\}$.
    If this is violated for $i\in S$, then raising $\effv_i$ to $\effv_i'$ cannot decrease $E$ (it can only increase $\sum_{s \in S} a_s(\effv)$) and cannot increase $\lambda$. 
    Similarly, if the assumption violated for $i\not\in S$, then lowering $\effv_i'$ to $\effv_i$ cannot decrease $E$ (it can only decrease $\sum_{s \in S} a_s(\effv')$) and cannot increase $\lambda$. 
    
    Now, suppose there exists some $i \in S$ such that $\frac {k\cdot{} g(\effv_i) }{ \sum_{t=s}^{n} g(\effv_t) } > 1$. 
    Then we can reduce $\effv_i$ so that $\frac {k\cdot{} g(\effv_i) }{ \sum_{t=s}^{n} g(\effv_t) } = 1$ since this doesn't change $E$ but potentially decreases $\lambda$. 
    Finally, suppose there exists some $i \in S$ such that $\frac{ k\cdot{} g(\effv_i') }{ \sum_{t=s}^{n} g(\effv_t') } > 1$. 
    Then consider lowering $\effv'_i$ so that $\frac{ k\cdot{} g(\effv_i') }{ \sum_{t=s}^{n} g(\effv_t') } = 1$ and then scaling $\effv''$ so that $\effv'_i$ has its original value. This does not change $E$ and potentially decreases $\lambda$.
    Therefore, we've successfully reduced to an instance in which the while loop never executes. 

    We now assume without loss of generality that the while loop never executes. The remaining argument follows closely from \cite{CIJ20_Multi_Category_Fairness}.

    Define $\alpha \coloneqq \sum_{s \in S} g(\effv_s)$ and $\beta \coloneqq \sum_{s \not\in S} g(\effv_s)$, and define $\alpha'$ and $\beta'$ analogously. 
    Note now that the while loop never executes, we have that for all $i$, $a_i(\effv) = g(\effv_i) / \sum_{i=1}^n g(\effv_s)$, and similarly for $a_i(\effv')$. 
    Therefore we can write $$E = \frac{\alpha}{\alpha + \beta} - \frac{\alpha'}{\alpha' + \beta'} = 1 - \frac{\beta}{\alpha + \beta} - \frac{\alpha'}{\alpha' + \beta'} .$$
    
    Let $R_\alpha \coloneqq \alpha/\alpha'$ and $R_\beta \coloneqq \beta'/\beta$. Note that $R_\alpha \leq g(\lambda)$ because for any $s \in S $, $\effv_s/\effv'_s \leq \lambda$ so $g(\effv_s)/g(\effv'_s) \leq g(\lambda)$. Similarly, $R_\beta \leq g(\lambda)$. Observe also that our expression for $E$ can be upper bounded by the case that these inequalities for $R_\alpha$ and $R_\beta$ are tight. 
    
    \begin{align*}
        E &\leq 1 - \frac{\alpha\cdot{} g(\lambda)}{\alpha\cdot{} g(\lambda) + \beta'} -  \frac{\alpha}{\alpha + \beta \cdot{}g(\lambda) } \\
        & = \frac{\alpha \beta' ( g(\lambda)^2 - 1 )}{(\alpha + \beta' g(\lambda)) (g(\lambda) \alpha + \beta') } \\
        & = \frac{\alpha \beta' ( g(\lambda)^2 - 1 )}{g(\lambda)\alpha^2 + g(\lambda) \beta'^2 + \alpha \beta' (g(\lambda)^2 + 1)} \\
        &\leq \frac{\alpha \beta' ( g(\lambda)^2 - 1 )}{2g(\lambda)\alpha\beta' + \alpha\beta' ( g(\lambda)^2 +1 )} \\
        &= \frac{g(\lambda)^2 - 1}{2g(\lambda) + g(\lambda)^2 +1 } \\
        &= \frac{g(\lambda) - 1 } {g(\lambda) + 1 }
    \end{align*}
    
    Finally, we observe that $\frac{g(\lambda) - 1 } {g(\lambda) + 1 } \leq f_\ell(\lambda)$, as desired: 
    $$E \leq \frac{\lambda^\ell - 1}{\lambda^\ell + 1}  =  1- 2(\lambda^\ell + 1)^{-1} \leq 1 - 2 (\lambda^\ell + \lambda^\ell)^{-1} = 1 - \lambda^{-\ell} \leq 1 - \lambda^{-2\ell} = f_\ell(\lambda)$$

\end{proof}
  
\begin{flushleft}
\textbf{\Cref{thm:PA_fairness}.}
    The \mpaname\ mechanism with parameter $g(x) = x^\ell$ satisfies \fullref{def:TV-value-stability} with respect to $f_\ell(\lambda)$. That is, for all pairs of effective value vectors $\effv, \effv'$, subsets of advertisers $S \subseteq [n]$, and slots $j$, 
    $$ \abs{\sum_{s \in S} P_{s,j}(\effv) - \sum_{s \in S} P_{s,j}(\effv')} \leq 2 f_\ell(\lambda)$$
\end{flushleft}

\begin{proof}
    Fix the parameter $\ell$, some subset of advertisers $S$, and the slot $j$. We have

    \begin{align*}
     \abs{\sum_{s \in S} P_s(\effv) - \sum_{s \in S} P_s(\effv')} &= \abs{ \sum_{s \in S}\left[ \kpah{j}_i(\effv) - \kpah{j-1}_i(\effv) \right] - \sum_{s \in S}\left[ \kpah{j}_i(\effv') - \kpah{j-1}_i(\effv') \right] } \\
    &= \abs{ \sum_{s \in S}[ \kpah{j}_i(\effv) - \kpah{j}_i(\effv') ] + \sum_{s \in S}[ \kpah{j-1}_i(\effv') - \kpah{j-1}_i(\effv)] } \\
    &\leq \abs{ \sum_{s \in S}\kpah{j}_i(\effv) - \sum_{s \in S}\kpah{j}_i(\effv') } + \abs{ \sum_{s \in S}\kpah{j-1}_i(\effv') - \sum_{s \in S}\kpah{j-1}_i(\effv) } \\
    &\leq f_\ell(\lambda) +  f_\ell(\lambda) \\
    &= 2f_\ell(\lambda)
\end{align*}

where the second to last step is by \Cref{lemma:k-unit-PA-fairness}.
\end{proof}

\begin{flushleft}
\textbf{\Cref{thm:pa-ordered-value-stability}.}
    \mpaname\ with parameter $\ell$ satisfies \hfairness\ with respect to $f_\ell(\lambda)$. That is, for every set of value and CTR vectors $v$, $v'$, $\userCTRVector$, $\userCTRVector'$ and $\slotCTRVector$, as well as for any advertiser $i$ and any decreasing vector $h$ with 
     $1 \geq h_{1} \geq \ldots \geq h_{k} \geq 0$:
 
$$\abs{ \sum_{j =1}^k h_{j} \left(\mpa_{i,j}-\mpa'_{i,j}\right) } \leq f_{\ell}(\lambda) \text{ where } \lambda \text{ is defined as } \max_{i\in[n]} \left( \max\left\{\frac{\alpha_iv_i}{\alpha'_iv'_i}, \frac{\alpha'_iv'_i}{\alpha_iv_i}\right\}\right)$$
where
$\mpa= \separableMechanism{v,\userCTRVector,\slotCTRVector}$ and $\mpa'= \separableMechanism{v',\userCTRVector',\slotCTRVector}$.
\end{flushleft}

\begin{proof}
    Fix some vectors $v, v', \alpha, \alpha',$ and $\beta$, and the corresponding allocation matrices $\mpa$ and $\mpa'$. Consider some advertiser $i$. We begin by using the definition of \mpaname\ and then rearranging terms. Note that we define $h_{k+1} \coloneqq 0$ for notational simplicity. 
    \begin{align*}
        \abs{ \sum_{j =1}^k h_{j} \left(\mpa_{i,j}-\mpa_{i,j}\right)} &= \abs{ \sum_{j =1}^k h_{j} \left((\kpah{j}_i(\effv) -\kpah{j-1}_i(\effv)) - (\kpah{j}_i(\effv') -\kpah{j-1}_i(\effv')) \right) } \\
        &= \abs{ \sum_{j =1}^k \left( h_{j} (\kpah{j}_i(\effv) -\kpah{j-1}_i(\effv)) - h_{j} (\kpah{j}_i(\effv') -\kpah{j-1}_i(\effv')) \right) } \\
        &= \abs{ \sum_{j =1}^k \left(\kpah{j}_i(\effv)-\kpah{j-1}_i(\effv')\right) \left(h_{j} - h_{j+1} \right) }
    \end{align*}
    Now, observe that because $h_1 \leq 1$ and the coefficients $(h_{j} - h_{j+1})$  telescope, the sum of these coefficients is at most $1$. Since the expression is a weighted sum over columns of the differences in allocation at that column, the expression is bounded by the maximum difference in any column. But because \mpaname\ satisfies total variation value stability (by \Cref{lemma:k-unit-pa-tv-value-stability}), this is bounded by $f_\ell(\lambda)$ for all subsets of advertisers, including the singleton $i$, as desired. 
    
    $$\abs{ \sum_{j =1}^k h_{j} \left(\mpa_{i,j}-\mpa'_{i,j}\right) } = \abs{ \sum_{j =1}^k \left(\kpah{j}_i(\effv)-\kpah{j-1}_i(\effv')\right) \left(h_{j} - h_{j+1} \right) } = \abs{ \max_j \left( \kpah{j}_i(\effv)-\kpah{j-1}_i(\effv') \right) } \leq f_\ell(\lambda)$$
\end{proof}

\subsection{Social Welfare}
\begin{flushleft}
\textbf{\Cref{lemma:k-unit-PA-SW}.}
  The \kpaname\ subroutine with parameter $\ell$ achieves an $\left( \frac{n-k}{n}(n-k)^{-1/\ell} + 1/n \right)$-approximation to the optimal social welfare for any instance with $n$ advertisers and $k$ slots. 
\end{flushleft}

   A key observation for the proof is that \mpaname\ achieves the desired social welfare approximation in the case that no advertiser is allocated an entire unit, shown below as \Cref{obs:k-unit-pa-sw-nofullunit}.

\begin{proof} 
    Consider the general case. Let $m$ be the number of advertisers allocated a full unit. Note that these will necessarily be $m$ highest bidders, and so the algorithm's allocation on these bidders is the same as under the optimal allocation. Hence, the algorithm's overall approximation ratio to the optimal is at least the approximation ratio that it achieves on the remaining $n-m$ advertisers. These advertisers receive the same allocations that they would in an instance with the first $m$ advertisers and first $m$ slots omitted. Therefore, we may assume we have $n-m$ advertisers and $k-m$ slots. By \Cref{obs:k-unit-pa-sw-nofullunit}, the mechanism on this instance achieves a social welfare approximation of at least $ \frac{n-k}{n-m}(n-k)^{-1/\ell} + 1/(n-m)$, which is at least our desired value of $\frac{n-k}{n}(n-k)^{-1/\ell} + 1/n$. This concludes the proof.
\end{proof} 

\begin{observation}\label{obs:k-unit-pa-sw-nofullunit}
   Let $(v,\alpha, \beta, n, k)$ be an instance in which the while loop in \Cref{PA-algorithm} never runs. Then we have $\swalg\geq \left( \frac{n-k}{n}(n-k)^{-1/\ell} + 1/n \right)\swopt$.
  \end{observation}
  \begin{proof}
  The proof of this lemma is very similar to the proof of social welfare guarantee in \cite{CIJ20_Multi_Category_Fairness}.
  
  Assume $v_1\geq \ldots \geq v_n$. Since we are assuming that the algorithm never enters the while loop, then the allocation to every advertiser $i$ will be $\frac{kg(v_i)}{\sum\limits_{j=1}^n g(v_j)}$ and so:
  $$\swalg=\sum\limits_{i=1}^n v_i\frac{k.v_i^\ell}{\sum\limits_{j=1}^n v_j^\ell}=k\frac{\sum\limits_{i=1}^n v_i^{\ell+1}}{\sum\limits_{i=1}^n v_i^\ell}$$
  Moreover, $\swopt=\sum\limits_{i=1}^k v_i$ because it allocates full units to the $k$ highest bidders. Now note that the algorithm is scale-free, so WLOG we may assume $\sum\limits_{i=1}^k = k$. Moreover, this means that $v_k\leq 1$, so $\vsum{k+1}{n}{\ell}\geq \vsum{k+1}{n}{\ell+1}$ because each $v_i$ for $i>k$ is at most $1$. Finally, note that $\vsum{1}{k}{\ell+1}\geq \vsum{1}{k}{\ell}$. (To see this, note that for $\ell=1$ the inequality holds by Cauchy–Schwarz inequality and equality happens at $v_i=1$ for all $i\in[k]$. By increasing $\ell$, $|v_i^\ell(v_i-1)|$ increases for $v_i\geq 1$ and decreases for $v_i<1$, so the inequality holds for any $\ell\geq 1$.)
  
  Now we want to show:
 \begin{equation*}
     \frac{\swalg}{\swopt}=\frac{k.\vsum{1}{n}{\ell+1}}{(\vsum{1}{n}{\ell})(\sum\limits_{i=1}^k v_i)}=\frac{\vsum{1}{n}{\ell+1}}{\vsum{1}{n}{\ell}}=\frac{\vsum{1}{k}{\ell+1}+\vsum{k+1}{n}{\ell+1}}{\vsum{1}{k}{\ell}+\vsum{k+1}{n}{\ell}}\geq \frac{1+\vsum{k+1}{n}{\ell+1}}{\vsum{k+1}{n}{\ell}}
 \end{equation*}
  To see this, let $A=\vsum{k+1}{n}{\ell+1}$, $B=\vsum{k+1}{n}{\ell}$, $X=\vsum{1}{k}{\ell+1}$, $Y=\vsum{1}{k}{\ell}$. We know that $X\geq Y\geq 1$ and $A\leq B$. We want to show that
  
      \begin{align*}
     \frac{X+A}{Y+B}\geq \frac{1+A}{1+B}&\\
     &\Longleftrightarrow(X+A)(1+B)\geq(Y+B)(1+A)\\
     &\Longleftrightarrow X+XB+A\geq Y+YA+B\\
     &\Longleftrightarrow (X_Y)+B(X-1)\geq A(Y-1)\\
\end{align*}
  Since $X-1\geq Y-1$ and $B\geq A$, the desired inequality holds. So we have:
  \begin{equation*}
      \frac{\swalg}{\swopt}\geq \frac{1+\vsum{k+1}{n}{\ell+1}}{1+\vsum{k+1}{n}{\ell}}
  \end{equation*}
  Now let $C=\vsum{k+1}{n}{\ell}$. For any $C$, we are interested in minimizing $\vsum{k+1}{n}{\ell+1}$. Now consider the following proposition from \cite{CIJ20_Multi_Category_Fairness}[Proposition A.5]:
  \begin{proposition}
  Consider $\Vec{x}\in (\mathbb{R}^{\geq 0})^n$ such that $||x||_\ell^\ell=C$. Then $||x||_{\ell+1}^{\ell+1}\geq n(\frac{C}{n})^{\frac{\ell+1}{\ell}}$.
  \end{proposition}
  Using this, we have:
  $$\vsum{k+1}{n}{\ell+1}\geq (n-k)(\frac{C}{n-k})^{\frac{\ell+1}{\ell}}$$
  Let $c=(\frac{C}{n-k})^{\frac{1}{\ell}}$. Since for all $i\geq k+1$, $v_i\leq 1$, then $0\leq C\leq n-k$. Then we have that $\frac{\swalg}{\swopt}$ is lower bounded by
  \begin{equation*}
      \frac{1+(n-k)c^{\ell+1}}{1+(n-k)c^\ell}=c+\frac{1-c}{1+(n-k)c^\ell}
  \end{equation*}
  
  Note that $0\leq c\leq 1$. The derivative of this expression with respect to $c$ is
 \begin{equation*}
     \frac{(n-k)c^{\ell-1}}{((n-k)c^\ell+1)^2}((n-k)c^{\ell+1}+c(\ell+1)-\ell)
 \end{equation*}
 The sign of the derivative is then specified by sign of $P(c)=(n-k)c^{\ell+1}+c(\ell+1)-\ell$. $P(c)$ is an increasing function of  $c$, $P(1)=n$ and $P(0)=-\ell$. Therefore, it has exactly one root $c^*\in(0,1)$. So the ratio $\frac{\swalg}{\swopt}$ is decreasing for $c<c^*$ and increasing for $c>c^*$. One can show that for appropriate ratios of $n$ and $k$, for $c'=(n-k)^{\frac{-1}{\ell}}$, $P(c')<0$ and so $c'<c^*$. Now observe that for $c\geq c'$,
 \begin{equation*}
     c+\frac{1-c}{1+(n-k)c^\ell}\geq c+\frac{1-c}{1+(n-k)}=\frac{c(n-k)}{n}+\frac{1}{n}
 \end{equation*}
  This bound increases with $c$, so plugging $c=c'$ gives the desired lower bound.
  \end{proof}

\begin{flushleft}
\textbf{\Cref{thm:PA_SW}.}
    The \mpaname\ mechanism with parameter $\ell$ achieves a  $\left( \frac{n-k}{n}(n-k)^{-1/\ell} + 1/n \right)$-approximation to the optimal social welfare for any instance with $n$ advertisers and $k$ slots. 
\end{flushleft}

\begin{proof}
    First, we consider the social welfare attained by the \spaname\ mechanism. 
    \begin{align*}
     \swalg &=\sum_{i =1}^n \sum_{j=1}^k \userCTR_i v_i \slotCTR_j M_{ij} \\ 
     &=\sum_{i=1}^n \sum_{j=1}^k \effv_i \slotCTR_j \left[\kpah{j}_i - \kpah{j-1}_i\right]\\ 
     &=\sum_{i=1}^n \sum_{j=1}^k \effv_i(\slotCTR_j-\slotCTR_{j+1}) \kpah{j}_i \ \ \text{ since } \slotCTR_{k+1} = 0 \text{ and } \kpah{0}_i= \zerovector \\
     &= \sum_{j=1}^k(\slotCTR_j-\slotCTR_{j+1}) \left(\sum_{i=1}^n \effv_i \kpah{j}_i \right)
\end{align*}

\Cref{lemma:k-unit-PA-SW} proves the approximation ratio of the $k$-unit PA mechanism. Observe that this ratio is decreasing in $k$. Therefore, for any $j$, $\left(\sum_{i=1}^n \effv_i \kpah{j}_i \right)$ is at least an $\eta = \left( \frac{n-k}{n}(n-k)^{-1/\ell} + 1/n \right)$ fraction of \opt. Therefore, we have

\begin{align*}
    \swalg 
    &= \sum_{j=1}^k(\slotCTR_j-\slotCTR_{j+1}) \left(\sum_{i=1}^n \effv_i \kipah{j}_i \right)\\
    &\geq \eta \sum_{j=1}^k (\slotCTR_j-\slotCTR_{j+1})\left(\effv_1 + \cdots + \effv_j\right) \\ 
    &= \eta \sum_{j=1}^k \effv_j\slotCTR_j \\
    &= \eta \, \swopt
\end{align*}
\end{proof}

\remove{
\section{Strong value stability}

\begin{definition}\label{def:strongly-value-stable}
[Strong value stability for position auctions] An allocation mechanism $\separableMechanism{\cdot}$ is $\valueStable$ with respect to function $f: [1,\infty]\rightarrow [0,1]$ if the following condition is satisfied for every set of value and CTR vectors $v$, $v'$, $\userCTRVector$, $\userCTRVector'$ and $\slotCTRVector$:
$$|\mipa_{i}-\mipa'_{i}|_1\leq 2f_{\ell}(\lambda) \text{ for all } i\in[n], \text{ where } \lambda \text{ is defined as } \max_{i\in[n]} \left( \max\left\{\frac{\alpha_iv_i}{\alpha'_iv'_i}, \frac{\alpha'_iv'_i}{\alpha_iv_i}\right\}\right)$$
where
$\mipa= \separableMechanism{v,\userCTRVector,\slotCTRVector}$ and $\mipa'= \separableMechanism{v',\userCTRVector',\slotCTRVector}$.
\end{definition}
\red{We have counter example that this guarantee does not hold. Do we want to keep this definition? Or just mention the counter example in appendix? The counter example is included in the appendix right now in appendix A}

\begin{flushleft}
\textbf{\Cref{ipa:strong}.} \textbf{Strong value stability} does not hold for generalized IPA mechanism.
\end{flushleft}
\begin{proof}
\red{TODO}
\end{proof}
}

\end{document}